# Article

# Visualizing dynamics of charges and strings in (2 + 1)D lattice gauge theories

Lattice gauge theories (LGTs)[1–4] can be used to understand a wide range of phenomena, from elementary particle scattering in high-energy physics to effective descriptions of many-body interactions in materials[5–7]. Studying dynamical properties of emergent phases can be challenging, as it requires solving many-body problems that are generally beyond perturbative limits[8–10]. Here we investigate the dynamics of local excitations in a $\mathbb{Z}_2$ LGT using a two-dimensional lattice of superconducting qubits. We first construct a simple variational circuit that prepares low-energy states that have a large overlap with the ground state; then we create charge excitations with local gates and simulate their quantum dynamics by means of a discretized time evolution. As the electric field coupling constant is increased, our measurements show signatures of transitioning from deconfined to confined dynamics. For confined excitations, the electric field induces a tension in the string connecting them. Our method allows us to experimentally image string dynamics in a (2+1)D LGT, from which we uncover two distinct regimes inside the confining phase: for weak confinement, the string fluctuates strongly in the transverse direction, whereas for strong confinement, transverse fluctuations are effectively frozen[11,12]. We also demonstrate a resonance condition at which dynamical string breaking is facilitated. Our LGT implementation on a quantum processor presents a new set of techniques for investigating emergent excitations and string dynamics.

Present models for fundamental forces are formulated as gauge theories. The common element of these theories is a local symmetry action and its corresponding gauge field that mediates interaction between matter particles[5]. Gauge theories are not limited to high-energy physics but can also capture emergent phenomena in condensed matter physics[6,7] and have seen applications in quantum information[13]. One of the earliest examples of the interplay between these research fields was the development of LGT, in which space is discretized to a lattice[1–4]. In particular, a motivation for introducing quantum LGTs was to describe a mechanism for confinement of quarks in quantum chromodynamics[2]. Within this framework, confined matter particles are the open ends of a string with finite tension. The discrete nature of LGTs has also been important in forming a framework for numerical calculations of equilibrium properties, for instance, using Monte Carlo or tensor-network-based methods[9].

Understanding the non-equilibrium dynamics of string excitations in LGTs is of fundamental importance in various disciplines, ranging from transport properties of the quark–gluon plasma to spectral properties in correlated quantum materials. However, theoretical approaches to this problem face substantial obstacles: non-equilibrium dynamics is beyond perturbative treatments, numerical methods based on Monte Carlo run into sign problems and tensor-network approaches work only as long as entanglement remains sufficiently low[8–10]. Quantum devices have been proposed as a viable alternative for the study of LGTs (refs. 14–20 for early works and reviews); their experimental implementations, on the other hand, have been limited to one spatial dimension or small scales, which limits the ability to examine string fluctuations[21–35]. Because conventional

LGT Hamiltonians have a constrained structure dictated by the local symmetry action, directly simulating their dynamics on quantum processors requires the ability to perform evolution generated by specific multibody local terms.

Here we realize a two-dimensional LGT on a superconducting quantum processor and use this platform to investigate and visualize the string dynamics. We consider an LGT in which the interaction between matter fields (filled circles in Fig. 1a), placed on the vertices of a square lattice, is mediated by $\mathbb{Z}_2$ gauge fields, located on the links that connect them[3] (green diamonds in Fig. 1a). This structure is a simplification of quantum electrodynamics, in which both space and the gauge group are discretized: space becomes a lattice and the $U(1)$ gauge group is discretized to $\mathbb{Z}_2$. We make use of the gauge redundancy to eliminate matter fields[36,37]. In the resulting 'matter-free' LGT, the motion and interaction of matter fields are captured by the $\mathbb{Z}_2$ gauge fields with the Hamiltonian:

$$\mathcal{H} = -J_E \sum_v A_v - J_M \sum_p B_p - h_E \sum_{\text{links}} Z_l - \lambda \sum_{\text{links}} X_l. \tag{1}$$

The vertex operators $A_v = \prod_{l \in v} Z_l$ are products of local $Z$ operators on link qubits emanating from a vertex $v$ and represent the electric charge (red or blue tiles in Fig. 1a). The plaquette operators $B_p = \prod_{l \in p} X_l$ are products of Pauli-X operators on link qubits encircling a plaquette and represent the presence or absence of magnetic flux (yellow or purple tiles). We consider vertex and plaquette operators of equal strength, which sets the unit of energy, $J_E = J_M = 1$. The $h_E$ terms denote an electric field on each link that creates magnetic flux excitations. The coupling terms $\lambda$ generate hopping of matter fields located at adjacent vertices, as mediated by a gauge field on their connecting link.

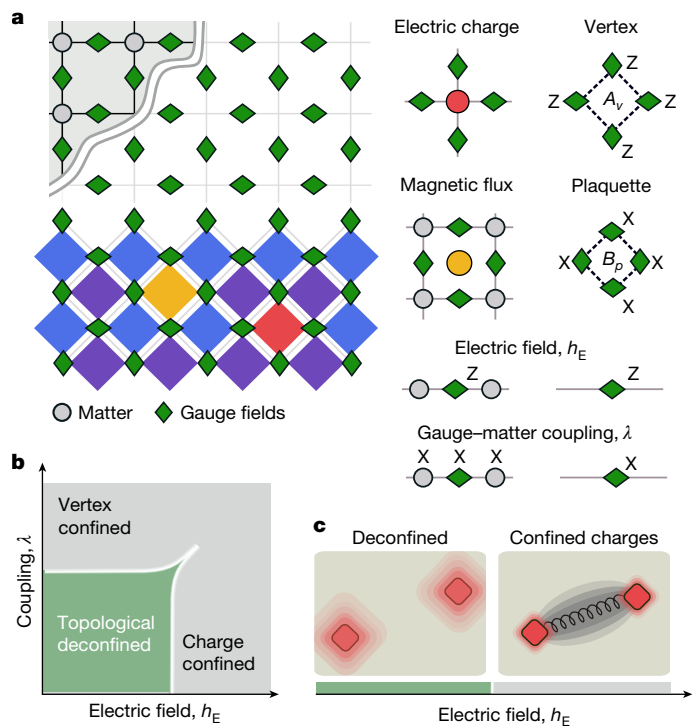

**Fig. 1 | An LGT and its phase diagram. a**, A full two-dimensional LGT (top left) can be realized by placing charged matter (grey circles) on vertices of a square lattice and gauge fields on the links between them (green diamonds). The local gauge structure can be used to eliminate the matter field and arrive at an effective theory involving gauge fields only (right). The presence/absence of charge excitations (red/blue) or magnetic fluxes (yellow/purple) is then sensed through the links. **b**, Zero-temperature phase diagram of the LGT in equation (1). **c**, In the deconfined phase, charges move freely. In the confined phase, charges oscillate around an equilibrium configuration. We can picture an elastic string connecting them that fluctuates in both longitudinal and transverse directions, limiting their motion.

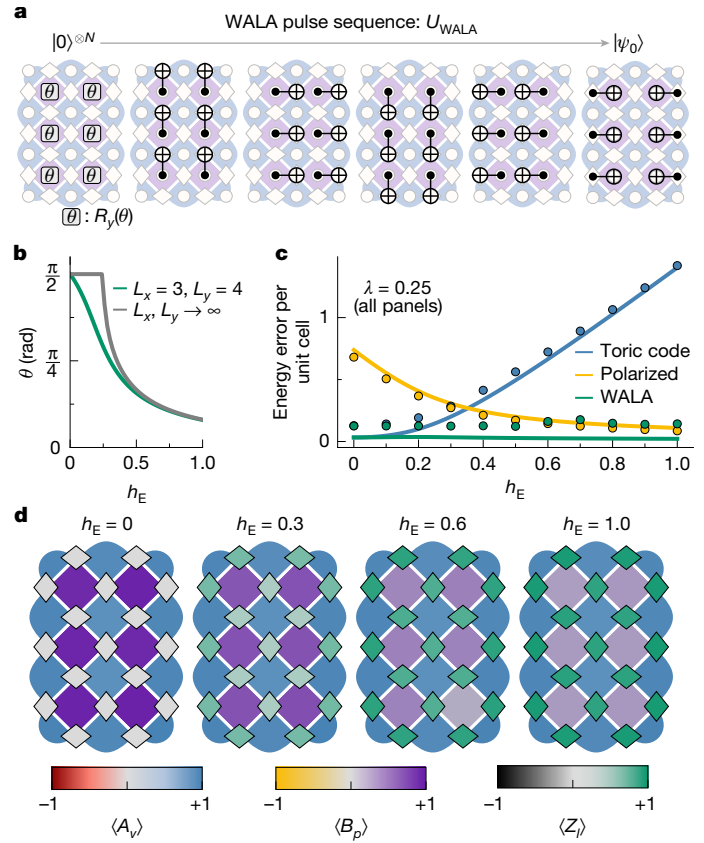

**Fig. 2 | WALA. a**, WALA gate sequence used for a two-dimensional grid of 35 qubits, consisting of 17 link qubits (diamonds) and 18 ancilla qubits (circles). The sequence begins with applying $R_y(\theta)$ to ancilla qubits of each plaquette, followed by applying C-NOT gates to qubit pairs, starting at the centre columns and moving outwards. **b**, Optimized $\theta$ angle used in WALA. The green curve is based on numerical calculations for a 35-qubit grid and the grey curve shows the thermodynamic limit. **c**, Energy error, compared with the exact ground state, of three ansatzes: (1) WALA (green); (2) toric code, $\theta = \pi/2$ (blue); and (3) product state, $|0\rangle^{\otimes N}$ (yellow), for $\lambda = 0.25$. Solid lines correspond to circuit simulations and filled circles are extracted from our experiment after readout error mitigation (Supplementary Information Section III.A). **d**, Experimentally measured expectation values of plaquette, vertex and Pauli-Z operators, for $\lambda = 0.25$ and $h_E \in \{0, 0.3, 0.6, 1.0\}$, from WALA. We post-select the measured data on the ancilla being in the expected $|0\rangle$ state to mitigate decoherence of the device for this and all other figures of the main text (Methods).

Since the foundational work of Fradkin and Shenker, it has been known that the zero-temperature phase diagram of $\mathcal{H}$ has two distinct phases[38–41] (Fig. 1b). One phase is the deconfined and topologically ordered phase that exists near $h_E, \lambda \approx 0$. The quantum phase transition along the $\lambda = 0$ line can be understood by a duality mapping to the transverse-field Ising model[1], in which domain walls of the Ising model correspond to closed strings in $\mathcal{H}$. For small but non-zero $\lambda$, the duality breaks down: we must include contributions to dynamics from open strings, which cannot correspond directly to domain walls. Crossing this transition into the confining phase leads to a condensation of magnetic excitations and confinement of electric excitations. The deconfinement to confinement transition can be seen in the non-equilibrium dynamics of a pair of charge excitations. In the deconfined case, the excitations move freely, whereas in the confined case, the string between them acquires a tension and restricts their motion (Fig. 1c).

For $h_E = \lambda = 0$, $\mathcal{H}$ reduces to the celebrated toric code Hamiltonian[13], which underlies several quantum computing error-correction codes. In that limit, all terms in $\mathcal{H}$ commute with each other, $[A_v, B_p] = 0 \; \forall v, p$; hence $\mathcal{H}$ is exactly solvable. The efficient preparation of the toric code ground state is well studied and can be achieved with circuits that scale linearly with the shorter dimension of the lattice[42–44]. In the limit $h_E, \lambda \gg 1$, the ground state is a product state of the qubits with all qubits pointing in the same direction, which can be prepared with single-qubit operations. A key aspect of $\mathcal{H}$ at the system sizes we study is the existence of an efficient algorithm to prepare states at energy densities low enough to resolve characteristic dynamics throughout the phase diagram. We make use of a variational ansatz based on a parameterization of the gauge structure used to generate the toric code wavefunction, which we call the weight-adjustable loop ansatz (WALA)[45,46] (Fig. 2a,b and Methods). To implement this ansatz, we use a grid of qubits with fourfold connectivity (diamonds) and ancilla qubits (circles; Fig. 2a) at the centre of each plaquette of the link qubits. All qubits begin in the $|0\rangle$ state. The state-preparation sequence starts with a single-qubit rotation $R_y(\theta)$ on each of the ancilla qubits at the centre of the plaquettes. The rest of the gate sequence does not have any adjustable parameters and is composed of controlled NOT (C-NOT) gates that generate entanglement between the qubits, starting with the centre columns of plaquettes and spreading to the edges of the lattice. The final C-NOT gate disentangles the ancilla qubits, returning them to the $|0\rangle$ state. We use a classical computer to find the optimal angle $\theta$ that minimizes the initial state energy, as a function of $h_E$ (Fig. 2b and of the excitations as a function of time for different electric fields $h_E$. Lines represent guides for the eye. All error bars in this work show standard deviation of the mean and are smaller than the markers when not visible. **e**, Spatial maps of the probability that an electric vertex $A_{i,j}$ is excited, conditioned on the electric vertex $A_{\nearrow}$ being excited at time $t = 3.5$: $P(i,j|\nearrow)$. The different columns correspond to different $A_{\nearrow}$, indicated by the boxed $\nearrow$ symbols. The unconditional probability that $A_{\nearrow}$ is excited is written below the $\nearrow$ symbol. The colour scale represents $P(i,j|\nearrow)$ for all $i$ and $j$. The top and bottom rows show results for $h_E = 0$ and 2.0, respectively. We implement dynamical decoupling and randomized compiling to mitigate control errors, as well as idle dephasing (Methods).

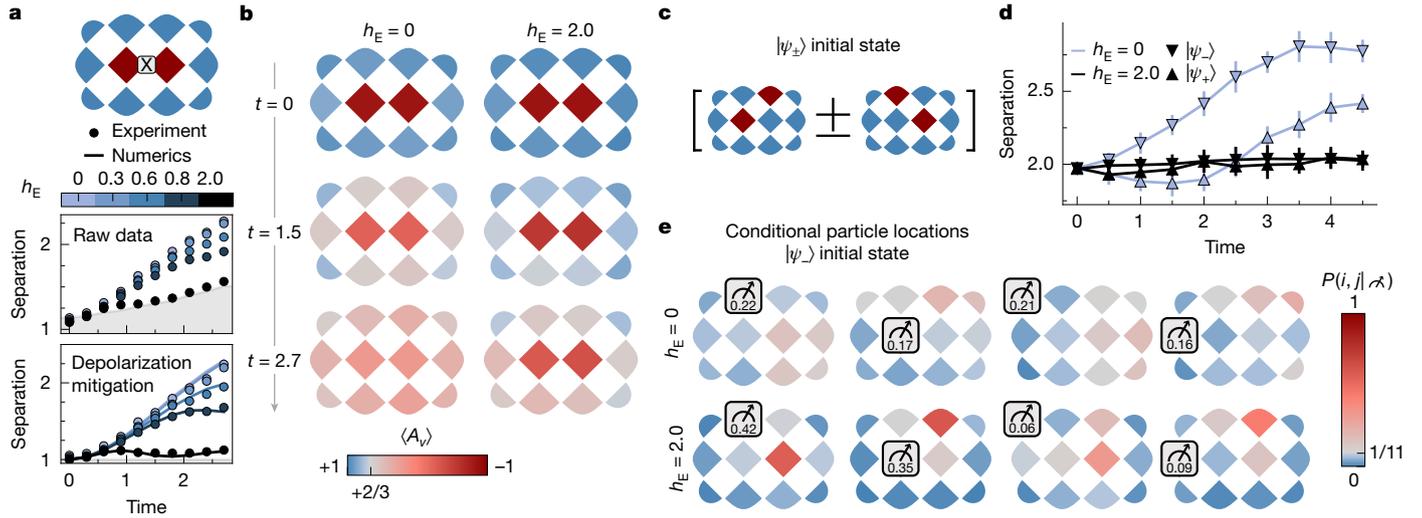

**Fig. 3 | Confinement of electric excitations. a**, A Pauli-X gate applied to the WALA initial state creates two electric excitations on adjacent vertices (red tiles). The coupling induces dynamics to the excitations and is set to $\lambda = 0.25$ for all data in this figure. After post-selecting bitstrings that correspond to two electric excitations, the separation of the excitations is monitored as a function of time for different electric fields $h_E$. The grey area is bounded by the separation measured when evolving under the pure toric code Hamiltonian ($\lambda = h_E = 0$). The lower panel shows the rescaled data, assuming a global depolarizing noise channel, and compares it with exact circuit simulation. **b**, Average density of electric excitations as measured by $\langle A_v \rangle$ for $h_E = 0$ (left) and $h_E = 2.0$ (right). **c**, Two superposition states in which the excitations interfere constructively ($|\psi_s\rangle$) and destructively ($|\psi_-\rangle$) at short distances, respectively. **d**, The separation

Supplementary Information Section II). The WALA circuit is equivalent to a mean-field ansatz for the dual Ising model and has the advantage of being very easy to optimize classically even for larger system sizes, in contrast to the actual ground state[45] (Methods). The resulting quantum state, $|\psi_0\rangle$, is then used as a low-energy-density initial state for simulating the dynamics. In classical simulations, the dynamics are hard to capture owing to the fast entanglement growth (Supplementary Information Section V).

In Fig. 2c, we show the energy error for $\lambda = 0.25$ as a function of $h_E$ using the optimized angles in the WALA (green markers/line). The energy error is small for all values of $h_E$ for the WALA initial state (Supplementary Information Section III.A). Preparing instead the toric code ground state ($\theta = \pi/2$) gives good overlap with the true ground state only for $h_E \ll 1$. Away from this limit, the energy error grows rapidly (blue markers/line). In the opposite limit, $h_E \gg 1$, the polarized state is the ground state of $\mathcal{H}$. Considering this ansatz yields acceptable energy errors for large values of $h_E$ (orange markers/line), but when reducing $h_E$, the energy error becomes large as well. The good performance of the WALA relies on the finite size of the system. In the thermodynamic limit, the WALA reduces to the toric code ground state for $h_E \in [0, h_{mf}]$, in which $h_{mf} = 0.25$ is the mean-field transition point (grey line in Fig. 2b). To characterize the WALA pulse sequence, we measure the expectation values of $A_v$ and $B_p$ and also Pauli-Z on individual qubits (diamonds in Fig. 2d). These local observables show changes of the various terms of $\mathcal{H}$. Owing to the form of our ansatz, all $\langle A_v \rangle$ (=1) and $\langle X_i \rangle$ (=0, not shown) terms remain constant as $\theta$ changes with increasing $h_E$. Variations in the energy density arise from the decrease of the magnetic parity values ($B_p$ terms) and the emergence of the Pauli-Z polarization values ($Z_i$ terms). The non-uniform $Z_i$ expectation values result from the distinct connectivity of the boundary qubits.

Having designed a circuit that approximates the ground state, we next study charge confinement by measuring the dynamics of a pair of electric excitations (Fig. 3). By using ancilla qubits at each vertex and plaquette centre, we are able to implement an efficient Suzuki–Trotter expansion of the time evolution operator generated by equation (1). Each time step has eight distinct layers, consisting of single-qubit rotations and controlled Z (CZ) gates, totalling 116 CZ gates per time step for the grid of 35 qubits (Methods). We prepare a pair of electric excitations on neighbouring sites in the centre of the system by applying a single Pauli-X on top of the WALA state (Fig. 3a). This operator creates a non-equilibrium initial state that has a notable overlap with the elementary excitations and allows us to examine the deconfinement–confinement dynamics of the charges. By measuring the average separation and the spatially resolved average position of these excitations as a function of time (Fig. 3a,b), we find that they show qualitatively distinct dynamical signatures as the electric field is tuned. Although for weak electric fields the excitations spread quickly across the whole system, at strong electric fields, the two charges stay together, as indicated by the small average separation; this observation constitutes a dynamical signature of charge confinement. Although the excitation separation increases for all values of $h_E$, we compare with the case when we evolve under the pure toric code Hamiltonian (grey region in Fig. 3a). The increase of separation in this case indicates that decoherence of the quantum state is pushing the system towards the maximally mixed state, which has an expected separation of 7/3. By also adding depolarization mitigation (Methods), we obtain quantitative agreement with numerical results and even reveal oscillations about an average separation that is much smaller than the system size when $h_E = 2.0$, indicative of a confining potential. These dynamical signatures support the onset of a confining potential near the Ising critical point, which—in the thermodynamic limit—is located between $h_E \approx 0.33$ and 0.34, depending on $\lambda$ as obtained from numerical ground-state studies[40,41,47,48].

To further accentuate the difference between confined and deconfined dynamics, we consider two other initial configurations, $|\psi_s\rangle$ and of interest ($Q_i$) may need to be mediated by means of swaps through an ancillary qubit ($Q_a$) and another gauge qubit ($Q_g$). **c**, Spatial maps of $\mathcal{S}_{ZZ}(t) = \text{Re}[\langle Z(t)Z(0)\rangle] \times \langle Z(0)\rangle$ for varying times and confining field $h_E$, at $\lambda = 0.25$ and $dt = 0.3$ (the same for all data in the figure). The extra qubits on either side used for state preparation are not shown. **d**, $\text{Re}[\langle Z(t)Z(0)\rangle]$ and $\mathcal{S}_{ZZ}(t)$ for qubits $Q_1$ and $Q_2$ in the centre-top and centre-bottom respectively, as labelled in panel **c**. Lines are guides for the eye. The grey regions in these plots correspond to the region limited by decoherence and are bounded by $|\langle Z(t)Z(0)\rangle|_{\lambda=h_E=0}$.

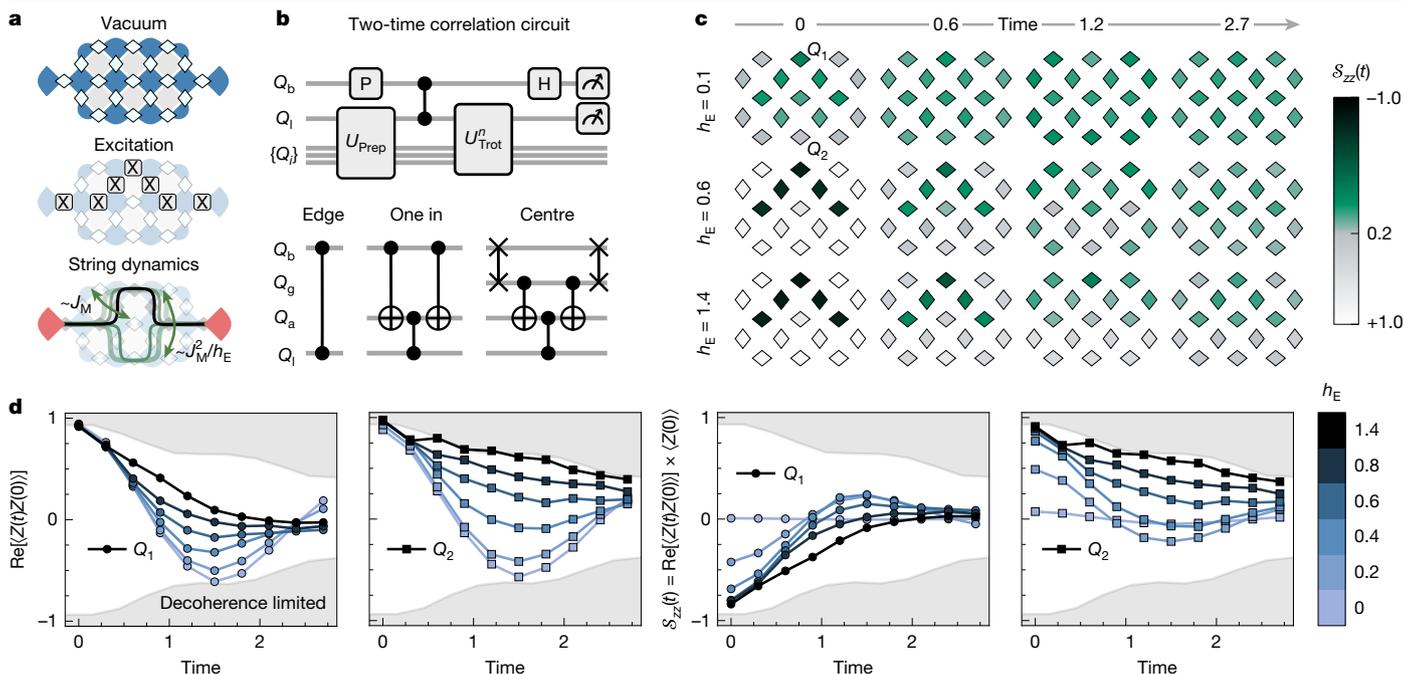

**Fig. 4 | Dynamics of the string connecting two spatially localized electric charges. a**, Schematic of the initial state preparation. Starting from the WALA initial state as vacuum, we create a pair of separated electric excitations by applying a string of X gates spanning from an extra qubit on the left (leftmost diamond) to one on the right (rightmost diamond). By not applying the local field terms of the time evolution on those two extra qubits, the excitations remain pinned, whereas the string itself can evolve dynamically. **b**, Circuit for measuring the unequal-time correlation function $\text{Re}[\langle Z(t)Z(0)\rangle]$. The P gate is $H(S^\dagger)^b$, in which $b = 0/1$ corresponds to measuring the real/imaginary part of $\langle Z(t)Z(0)\rangle$. The CZ gate entangling the auxiliary qubit ($Q_b$) to the gauge qubit

$|\psi_-\rangle$ (Fig. 3c). These are positive and negative superpositions of a pair of excitations at a lattice distance of two. The intuition for choosing these initial states comes from approximations to different angular momentum eigenstates. Their dynamics can be understood from quantum interference: at short times, hopping of electric excitations that brings the pair closer together interferes constructively for $|\psi_+\rangle$ and destructively for $|\psi_-\rangle$. In the deconfined phase, this leads to the excitations initially moving closer together for $|\psi_+\rangle$ and further apart for $|\psi_-\rangle$, as observed in Fig. 3d for $h_E = 0$. By contrast, when $h_E = 2.0$, the string tension dominates and the excitations remain close to their initial separation for both initial states. The excitation separation initialized in this state has the added feature of being robust against Trotter error, allowing us to increase the Trotter step from $dt = 0.3$ to $dt = 0.5$ and reach later times (Supplementary Information Section IV.C), increasing the signal strength. The confinement signatures observed in excitation separations can be further corroborated by analysing the probability of finding an excitation at a given site, conditioned on measuring another excitation somewhere else in the lattice (Fig. 3e). The data shown are for a fixed time $t = 3.5$ for the $|\psi_-\rangle$ configuration (see Methods for $|\psi_+\rangle$). For $h_E = 0$, the probabilities are spread across the system, with a higher probability of the charges being found further apart than their initial separation. For $h_E = 2.0$, there is only a substantial probability observed at separations 1, 2 and 3, indicating that the excitations tend to stay close to their initial position or hop together in a correlated fashion and demonstrating confinement of pairs of electric charges.

In the confined regime, electric charges are located at the ends of an elastic string, which—in our two-dimensional setting—can vibrate transversely, akin to a violin string. We generate the string by applying X gates that traverse the system on top of the WALA circuit from an auxiliary qubit on the left to another one on the right (Fig. 4a). By performing a Trotterized time evolution, which excludes field terms on these extra qubits, the so-created electric charges will remain pinned at the edges, whereas the string itself can evolve dynamically. To investigate the vibrational dynamics of the string, we measure a two-time correlator in the Z basis:

$$\mathcal{S}_{ZZ}(t) = \text{Re}[\langle Z(t)Z(0)\rangle] \times \langle Z(0)\rangle \quad (2)$$

for each qubit. We measure $\mathcal{S}_{ZZ}(t)$ using a Hadamard test with an auxiliary circuit (Fig. 4b and Methods). This correlation function is a product of two terms. The first term is sensitive to whether the presence of the string has changed compared with its initial value at time zero, that is, it captures the stiffness of the string. The second term measures whether a string has been created on top of the WALA sequence initially, which is only possible in the confined regime. The combined correlation function, $\mathcal{S}_{ZZ}(t)$, allows us to determine the string dynamics. Note that, although for $\lambda = 0$ strings correspond to Ising domain walls, for finite fields $\lambda \neq 0$, there exists no direct mapping to domain walls.

Our measurements of $\mathcal{S}_{ZZ}(t)$ reveal three distinct regimes (Fig. 4c,d). (1) In the deconfined phase, $h_E = 0.1$, applying an X-string on top of the WALA sequence does not create a string excitation. The correlation function $\text{Re}[\langle Z(t)Z(0)\rangle]$ experiences coherent oscillations from the energy difference between the WALA state and the first excited state, whereas the small expectation value of $\langle Z(0)\rangle$ suppresses $\mathcal{S}_{ZZ}(t)$. (2) In the intermediate electric field regime, $h_E = 0.6$, the dynamics are already confining but the string tension is not too large. Thus, changes of the string length from higher-order dynamical processes are energetically accessible. Our measurements show a clear initial string along the path pair (Fig. 5a). When, by contrast, this process occurs on a string, there is a string-energy gain that competes with the energy cost of the pair creation.

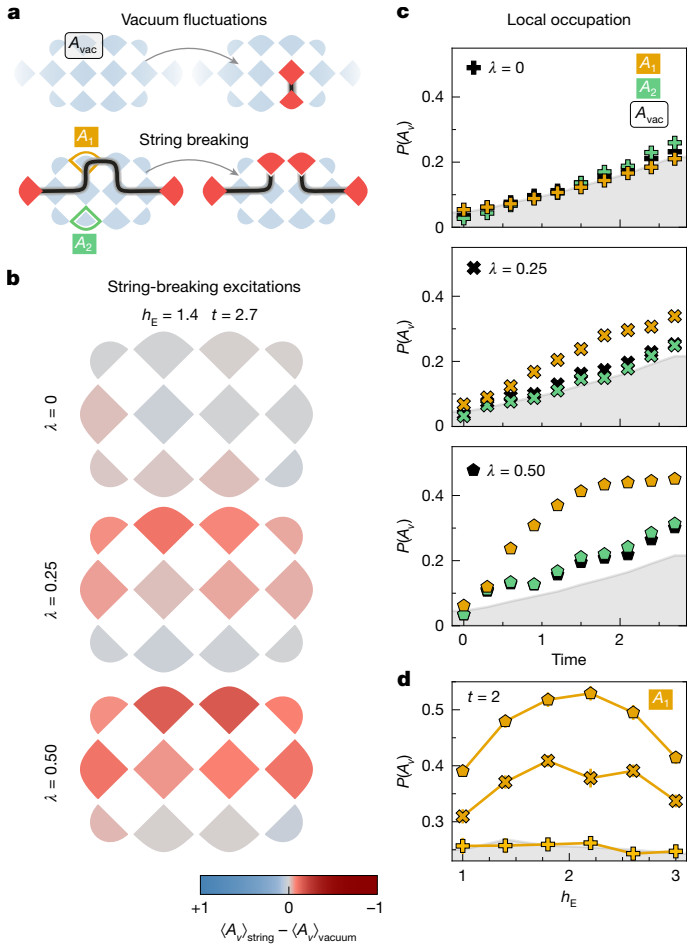

**Fig. 5 | String breaking. a**, Schematics for pair creation from vacuum fluctuation and string breaking. **b**, Difference in the charge excitation values in the presence and absence of the string $\langle A_v \rangle_{\text{string}} - \langle A_v \rangle_{\text{vacuum}}$ for $\lambda \in \{0, 0.25, 0.50\}$ at $h_E = 1.4$ and $t = 2.7$, with $dt = 0.3$. **c**, Probability of a vertex excitation $P(A_v)$ on three distinct vertices $A_1$ (gold), $A_2$ (green) and $A_{\text{vac}}$ (black) for $\lambda \in \{0, 0.25, 0.50\}$ and $h_E = 1.4$. The grey 'decoherence-limited' region is defined by the average of $P(A_v)$ over all vertices having evolved the initial state with the X string for $\lambda = 0$, $h_E = 1.4$. **d**, Dependence of $P(A_v)$ on $h_E$, acquired at $t = 2$ ($dt = 0.2$), for $\lambda = 0$ (pluses), $\lambda = 0.25$ (crosses) and $\lambda = 0.50$ (pentagons).

To investigate string breaking in our experiment, we first measure the electric excitations through $\langle A_v \rangle$ in the presence and absence of an initial string excitation in the confined regime $h_E = 1.4$. In Fig. 5b, we show the difference of $\langle A_v \rangle$ for these two initial states at time $t = 2.7$ for different strengths of string breaking set by the coupling $\lambda$. When $\lambda = 0$, the charge density in the presence of the string is comparable with the one in the absence of the string, $\langle A_v \rangle_{\text{string}} - \langle A_v \rangle_{\text{vacuum}} \simeq 0$. However, for finite $\lambda$ values, the time-evolved string initial state has considerably more charges. We track the dynamics of $\langle A_v \rangle$ for electric excitations at the top and at the bottom of the system. As demonstrated, for this confinement field $h_E$, the string stays mainly at the top qubits within the accessible timescales. For $\lambda = 0$, both vertex operators $A_1$ on the top and $A_2$ at the bottom show the same trend as the vacuum state $A_{\text{vac}}$ (Fig. 5c), which can be understood from decoherence of the device (grey region). However, as $\lambda$ is increased to 0.25 and 0.50, the electric charge $A_1$ on the top side, at which the string has been created, shows a much higher number of excitations compared with $A_2$, which remains indistinguishable from the vacuum. This differential measurement is evidence for excitation creation from string breaking.

In the lowest-order string-breaking process, the coupling reduces the length of the string by one, leading to an energy gain of $2h_E$ and, at the same time, two electric excitations are created with a cost of $4J_E$. Therefore, we predict that this energy trade-off could enhance the probability of string breaking near $h_E = 2J_E$. We measure the probability of electric charge creation $P(A_v)$ on $A_1$ as a function of $h_E$ (Fig. 5d). For finite couplings, we observe a maximum charge creation in the vicinity of $h_E \approx 2$, demonstrating that string breaking is facilitated at the resonance condition.

In this work, we imaged the dynamics of deconfined and confined excitations in a (2+1)D $\mathbb{Z}_2$ LGT and measured the vibrations of the string connecting them. Our work demonstrates the potential for quantum processors to study the dynamics of emergent excitations in correlated quantum matter, which are prohibitively hard to predict theoretically owing to their non-perturbative nature. The dynamical observables measured in this work are related to those measured in more conventional scattering experiments but examine complementary regimes; our approach provides insight into the microscopic nature of non-equilibrium dynamics of quasiparticles, as compared with the measurement of their spectral properties. Our space-resolved and time-resolved measurements provide a new visualization approach for characterizing the dynamics of interacting emergent excitations.

## Online content

Any methods, additional references, Nature Portfolio reporting summaries, source data, extended data, supplementary information, acknowledgements, peer review information; details of author contributions and competing interests; and statements of data and code availability are available at https://doi.org/10.1038/s41586-025-08999-9.

we prepared. Already after a short temporal evolution, the $\mathcal{S}_{ZZ}(t)$ correlations of the qubits both at the bottom and at the top of the grid quickly decay to zero and the string is equally likely found on either side of the system. In this regime, the string is floppy and fluctuates strongly, even though charges remain confined[11,12]. (3) Deep in the confined regime, for large $h_E = 1.4$, we see dynamics of the initial bump at the top of the grid but very little probability that the string moves to the bottom on the timescales of the simulation. This can be understood from the large string tension in the deeply confined regime, which suppresses the flopping to the other side within experimentally accessible timescales. However, the string is still moving freely around the top qubits despite the large string tension. These length-scale-preserving moves result from the lattice discretization of our LGT, as the plaquette terms $B_p$ of the Hamiltonian can deform the string without changing its length.

Having visualized the vibrations of the string connecting two electric charges in the confined regime, we now investigate string-breaking and pair-creation dynamics. The coupling $\lambda$ can dynamically create pairs of electric excitations. When this process occurs in the ground state, the energy is increased as a result of the cost of creating an excitation

# Article

T. A. Cochran[1,2,13], B. Jobst[3,4,13], E. Rosenberg[1,13], Y. D. Lensky[1,13], G. Gyawali[1,5,6,13], N. Eassa[1,7], M. Will[3,4], A. Szasz[1], D. Abanin[1], R. Acharya[1], L. Aghababaie Beni[1], T. I. Andersen[1], M. Ansmann[1], F. Arute[1], K. Arya[1], A. Asfaw[1], A. Atalaya[1], R. Babbush[1], B. Ballard[1], J. C. Bardin[1,8], B. Bengtsson[1], A. Bilmes[1], A. Bourassa[1], J. Bovaird[1], M. Broughton[1], D. A. Browne[1], B. Buchea[1], B. B. Buckley[1], T. Burger[1], B. Burkett[1], N. Bushnell[1], A. Cabrera[1], J. Campero[1], H.-S. Chang[1], Z. Chen[1], B. Chiaro[1], J. Claes[1], A. Y. Cleland[1], J. Cogan[1], R. Collins[1], P. Conner[1], W. Courtney[1], A. L. Crook[1], B. Curtin[1], S. Das[1], S. Demura[1], L. De Lorenzo[1], A. Di Paolo[1], P. Donohoe[1], I. Drozdov[1,9], A. Dunsworth[1], A. Eickbusch[1], A. Moshe Elbag[1], M. Elzouka[1], C. Erickson[1], V. S. Ferreira[1], L. Flores Burgos[1], E. Forati[1], A. G. Fowler[1], B. Foxen[1], S. Ganjam[1], R. Gasca[1], É. Genois[1], W. Giang[1], D. Gilboa[1], R. Gosula[1], A. Grajales Dau[1], D. Graumann[1], A. Greene[1], J. A. Gross[1], S. Habegger[1], M. Hansen[1], M. P. Harrigan[1], S. D. Harrington[1], P. Heu[1], O. Higgott[1], J. Hilton[1], H.-Y. Huang[1], A. Huff[1], W. Huggins[1], E. Jeffrey[1], Z. Jiang[1], C. Jones[1], C. Joshi[1], P. Juhas[1], D. Kafri[1], H. Kang[1], A. H. Karamlou[1], K. Kechedzhi[1], T. Khaire[1], T. Khattar[1], M. Khezri[1], S. Kim[1], P. Klimov[1], B. Kobrin[1], A. Korotkov[1,10], F. Kostritsa[1], J. Kreikebaum[1], V. Kurilovich[1], D. Landhuis[1], T. Lange-Dei[1], B. Langley[1], K.-M. Lau[1], J. Ledford[1], K. Lee[1], B. Lester[1], L. Le Guevel[1], W. Li[1], A. T. Lill[1], W. Livingston[1], A. Locharla[1], D. Lundahl[1], A. Lunt[1], S. Madhuk[1], A. Maloney[1], S. Mandrà[1], L. Martin[1], O. Martin[1], C. Maxfield[1], J. McClean[1], M. McEwen[1], S. Meeks[1], A. Megrant[1], K. Miao[1], R. Molavi[1], S. Molina[1], S. Montazeri[1], R. Movassagh[1], C. Neill[1], M. Newman[1], A. Nguyen[1], M. Nguyen[1], C.-H. Ni[1], K. Ottosson[1], A. Pizzuto[1], R. Potter[1], O. Pritchard[1], C. Quintana[1], G. Ramachandran[1], M. Reagor[1], D. Rhodes[1], G. Roberts[1], K. Sankaragomathi[1], K. Satzinger[1], H. Schurkus[1], M. Shearn[1], A. Shorter[1], N. Shutty[1], V. Shvarts[1], V. Sivak[1], S. Small[1], W. C. Smith[1], S. Springer[1], G. Sterling[1], J. Suchard[1], A. Sztein[1], D. Thor[1], M. Torunbalci[1], A. Vaishnav[1], J. Vargas[1], S. Vdovichev[1], G. Vidal[1], C. Vollgraff Heidweiller[1], S. Waltman[1], S. X. Wang[1], B. Ware[1], T. White[1], K. Wong[1], B. W. K. Woo[1], C. Xing[1], Z. Jamie Yao[1], P. Yeh[1], B. Ying[1], J. Yoo[1], N. Yosri[1], G. Young[1], A. Zalcman[1], Y. Zhang[1], N. Zhu[1], N. Zobrist[1], S. Boixo[1], J. Kelly[1], E. Lucero[1], Y. Chen[1], V. Smelyanskiy[1], H. Neven[1], A. Gammon-Smith[11,12], F. Pollmann[3,4 ✉], M. Knap[3,4 ✉] & P. Roushan[1 ✉]

[1]Google Research, Mountain View, CA, USA. [2]Department of Physics, Princeton University, Princeton, NJ, USA. [3]Department of Physics, TUM School of Natural Sciences, Technical University of Munich, Garching, Germany. [4]Munich Center for Quantum Science and Technology (MCQST), Munich, Germany. [5]Department of Physics, Cornell University, Ithaca, NY, USA. [6]Laboratory of Atomic and Solid State Physics, Cornell University, Ithaca, NY, USA. [7]Department of Physics and Astronomy, Purdue University, West Lafayette, IN, USA. [8]Department of Electrical and Computer Engineering, University of Massachusetts, Amherst, MA, USA. [9]Department of Physics, University of Connecticut, Storrs, CT, USA. [10]Department of Electrical and Computer Engineering, University of California, Riverside, Riverside, CA, USA. [11]School of Physics and Astronomy, University of Nottingham, Nottingham, UK. [12]Centre for the Mathematics and Theoretical Physics of Quantum Non-Equilibrium Systems (CQNE), University of Nottingham, Nottingham, UK. [13]These authors contributed equally: T. A. Cochran, B. Jobst, E. Rosenberg, Y. D. Lensky, G. Gyawali. ✉e-mail: frank.pollmann@tum.de; michael.knap@ph.tum.de; pedramr@google.com

## Methods

### Experimental techniques and device characterization

**Gate implementation.** All experiments in this work can be carried out on a grid of 45 qubits with square connectivity and were implemented on a 72-qubit Google Sycamore processor, as used in ref. 49 (Extended Data Fig. 1). Dominant errors come from CZ entangling gates[50] and final readout[51]. Qubit, coupler and readout parameters are optimized using the Snake Optimizer[52,53]. A smaller contribution to the total error comes from the single-qubit microwave gates. Gates are calibrated using Google's Optimus calibration tools[54,55].

To mitigate the effects of coherent noise, we implement randomized compiling[56–58]. For all observables, we average over 30 compiling instances of randomly chosen single-qubit Pauli gates sandwiching each CZ gate, such that the resulting ideal unitary is unchanged. We record approximately 400 shots per instance after post-selection. All sequential single-qubit gates are then combined into a single phased-X Z gate, such that the structure of the circuit is always alternating single-qubit and two-qubit gate layers. The use of randomized compiling to convert coherent errors to incoherent ones also supports our choice to apply simply simple depolarizing noise mitigation to compare experimental results with numerical simulations.

During the projective state preparation and Hadamard test experiments presented in this work, there are ancilla qubits that must sit unchanged while the rest of the system undergoes up to ten Trotter cycles, which constitute 80 single-qubit layers and 80 two-qubit layers. These long evolution times place a strict constraint on the ancilla qubit to remain highly coherent. Therefore, it is important to mitigate ancilla dephasing to ensure the highest fidelity experiments. To this end, we implement dynamical decoupling whenever a qubit would otherwise be idle. Our approach is to use a simple echo sequence of X gates during each single-qubit gate layer[42,59].

In implementing circuits formulated in terms of non-native C-NOT and SWAP gates, each C-NOT gate is converted to a CZ gate, sandwiched by Hadamard gates on the target qubit. The extra Hadamard gates then get combined with other sequential single-qubit gates into a single phased-X Z gate. SWAP gates, used to initialize central qubits for the Hadamard test in Fig. 4, are broken into three C-NOT gates.

**Suzuki–Trotter circuit.** In implementing the WALA state and time evolution under $\mathcal{H}$, the fourfold connectivity of Google's Sycamore quantum processor allows for a configuration without ancilla qubits[42] or with an ancilla qubit at the centre of each vertex and plaquette[49]. Although this configuration allows for a denser packing of vertex sites, it complicates the Trotter evolution because all plaquette and vertex terms cannot be executed in parallel, increasing errors from $T_1$ and $T_2$ decay. Instead, we use the configuration with ancilla qubits. The grid of qubits and the corresponding vertices and plaquettes are shown in Extended Data Fig. 2. The diamonds represent physical qubits on the gauge sites and the circles indicate ancilla qubits. The blue/purple tiles correspond to vertices/plaquettes, respectively.

To carry out the time evolution, we developed a circuit to implement the first-order Suzuki–Trotter expansion (Trotter errors discussed in Supplementary Information Section IV.C). The first operation of the Trotter circuit consists of a local field unitary operator:

$$U_{\text{Fields}} = \exp\left[-i\left(-\lambda \sum_l X_l - h_E \sum_l Z_l\right)dt\right] \tag{3}$$

which can be implemented by a single phased-X Z gate on each physical gauge qubit. The second operator of the Trotter circuit involves the vertex and plaquette terms:

$$U_{\text{Plaquettes}} = \exp\left[-i\left(-J_E \sum_v A_v - J_M \sum_p B_p\right)dt\right] \tag{4}$$

which can be implemented in parallel for all vertices and plaquettes in eight entangling layers. With four layers of entangling gates, the commuting $A_v/B_p$ operators are transformed into single-qubit operators on the ancilla qubits, which are then rotated by an angle $-2J_E dt/-2J_M dt$ about the Z axis to invoke the time evolution. The transformation of the vertex and plaquette operators is then reversed with another four layers of entangling gates, which returns the state to the physical qubits and disentangles the ancilla qubits. This algorithm gives a gate count of $16L_xL_y - 12(L_x + L_y) + 8$ per Trotter cycle for a rectangular grid with $L_x \times L_y$ vertex sites (116 entangling gates per Trotter cycle for our experimental set-up in Extended Data Fig. 2). Executing state preparation and nine Trotter steps, our experiment uses about 1,075 total entangling gates. Because we are focusing on local observables, we measure appreciable signals in these circuits, whose depth is slightly larger than state-of-the-art random circuit sampling experiments, which focus on cross-entropy benchmarking[60].

**Post-selection.** Decoherence is unavoidable on noisy intermediate-scale quantum (NISQ) processors. A common technique to deal with decoherence, which does not depend on any detailed knowledge of the device error model, is post-selection. Any observable that is known to be conserved by the quantum circuit is a good candidate for post-selection criteria, as long as they can be measured concurrently with the final observable of physical interest. In our case, we extract expectation values by measuring the physical qubits. However, we also measure the ancilla qubits concurrently, which would all remain in the $|0\rangle$ state if no errors occurred (Extended Data Fig. 3). This is because circuits for state preparation and Trotterization entangle the ancilla qubits with the physical qubits to carry out the quantum operation, but always disentangle the ancilla and ideally return it to its original state. However, errors during a Trotter step can quickly propagate across the chip. Our measurements show increasing numbers of these errors as the number of Trotter steps increases (Extended Data Fig. 3a,b). Therefore, to mitigate these errors, we post-select all of our measurements presented in the main text and supplement such that all ancilla qubits are in the $|0\rangle$ state.

Even after this first round of post-selection, residual experimental errors, Trotter errors and deviation from perfect overlap between the $A_v$ operators and dressed physical particle operators contribute to experimental measurements of a different number of excitations than initialized. Although the number of vertex parity violations is not an exactly conserved quantity, it is expected to be an approximate one. Therefore, in measurements that scrutinize the properties of a set number of electric excitations, we post-select shots that have the same number of vertex parity flips as the initialized state to further mitigate experimental errors, while also eliminating some of the spurious effects of the Trotter error (Fig. 3a,d,e, Extended Data Figs. 4 and 8 and Supplementary Figs. 3, 4 and 9). An added benefit from post-selecting on the two-charge sector is the ability to unambiguously assign the distance between two charges (Fig. 3a,d).

Although these post-selection techniques increase the accuracy of our results, they come at an exponential cost (Extended Data Fig. 3c). We find that, to obtain a constant number of post-selected shots after applying both ancilla qubit and excitation number post-selection on our standard grid of 17 physical qubits and 18 ancilla qubits, the number of shots taken on the hardware scales as $N_{\text{shots}} \approx 2.5^n$, for $n$ Trotter steps. This procedure reduces the rate that observables of a stationary state trend towards the maximally mixed value (Extended Data Fig. 3d).

**Global depolarizing channel mitigation.** Taking bitstrings directly from the experimental measurements and computing the various observables in this work results in deviations from expectations. For example, results on a 2 × 3 vertex system are shown in Extended Data Fig. 4a. Although there seems to be clear separation between the $h_E = 0$ data points (red) and the $h_E = 2.25$ data points (blue), the excitations seem to be drifting apart quickly for all values of $h_E$.

To estimate the effect of device noise, we perform numerical simulations with a local two-qubit depolarizing noise channel following every CZ gate in our circuit. We estimate the depolarizing probability to be 0.7%, corresponding to the mean error rate attained from cross-entropy benchmarking. The results are shown by the dashed lines in Extended Data Fig. 4a. We see that the local depolarizing error model captures the trend of the data very well for all values of $h_E$.

To simplify the interpretation and analysis, we consider the possibility that a global depolarizing channel may capture the behaviour of our data phenomenologically[61–64]. To determine the correct error probability to mitigate, we consider Trotterized Hamiltonian evolution under parameters that leave the observable unchanged. For example, when measuring the vertex parity values, time evolution with $\lambda = 0$ results in no dynamics in the ideal case, because all of the vertex operators commute with the Hamiltonian. Then, by monitoring the measured charge separation as a function of Trotter cycle, the expectation value drifts from the expectation for the initial state to that of the depolarized state (Extended Data Fig. 4b). The effective depolarizing probability, $p_{eff}$, can be extracted for each cycle:

$$p_{eff} = \frac{\langle \hat{\mathcal{O}} \rangle_{measured} - \mathcal{O}_{initial}}{\mathcal{O}_{depolarized} - \mathcal{O}_{initial}} \tag{5}$$

with $\langle \hat{\mathcal{O}} \rangle_{measured}$ being the measured expectation value, $\mathcal{O}_{initial}$ being the initial value of the observable, set by the state preparation, and $\mathcal{O}_{depolarized}$ being the expectation value in the completely depolarized state, that is, the maximally mixed state. Such a $p_{eff}$ for the data shown in Extended Data Fig. 4a is shown in Extended Data Fig. 4c. The solid lines in Extended Data Fig. 4a correspond to the application of the global depolarizing channel with the effective depolarizing probability in Extended Data Fig. 4c. The agreement between experiment and the global depolarizing model is reasonable.

To compare experimental measurements of an observable $\hat{\mathcal{O}}$ with noiseless simulations, we use this global depolarizing model to rescale the experimental data:

$$\langle \hat{\mathcal{O}} \rangle_{rescaled} = \frac{\langle \hat{\mathcal{O}} \rangle_{measured} - p_{eff} \mathcal{O}_{depolarized}}{1 - p_{eff}} \tag{6}$$

Although our experimental data innately show robust signatures of confinement and string dynamics without any rescaling, to compare with numerical simulations, we show rescaled data (always alongside its unscaled counterpart) in Fig. 3a, Extended Data Figs. 9 and 10c and Supplementary Figs. 2, 3, 4c, 5d, 6, 7 and 8.

**Two-time Pauli string correlator Hadamard test.** In this work, we present two distinct two-time Pauli string correlators of the form $\langle \psi | Z_l(t) Z_l(t_0) | \psi \rangle$ (Fig. 4) and $\langle \psi | (X_{Q_1} X_{Q_2} ... X_{Q_j})(t) X_{Q_1}(t_0) | \psi \rangle$ (Supplementary Information Section III.C.2). Theoretical works have outlined schemes for measuring such quantities with quantum circuits[65,66] and, recently, experiments have used Hadamard tests with two controlled operations to measure two-time correlators[67]. Because a generic controlled-$P$ operation, in which $P$ is an arbitrary Pauli string, is not necessarily native to our quantum hardware, we measure correlation functions with a version of the Hadamard test with only a single controlled operation, $C\text{-}A$ at time $t_0$ (ref. 68). In fact, because both of these correlators are of the form $\langle \psi | B(t) A(t_0) | \psi \rangle$ with simple operators $A = Z_l$ and $A = X_{Q_l}$, respectively, the implementation is straightforward and only requires C-NOT and CZ gates as controlled operations.

To measure the two-time Pauli string correlator, $\mathcal{C}(A(t_0), B(t)) = \langle \psi | B(t) A(t_0) | \psi \rangle$, we first prepare the ancilla in the state:

$$|\eta(\vartheta, \phi)\rangle = \cos\frac{\vartheta}{2} |0\rangle + \sin\frac{\vartheta}{2} e^{i\phi} |1\rangle \tag{7}$$

using an arbitrary single-qubit gate. The controlled operator we want to apply to the system to measure $\mathcal{C}$ is $C\text{-}A$. Thus, having the grid of qubits starting in state $|\psi\rangle$, we apply $C\text{-}A$ to the system, controlled by the ancilla, and have the following resulting state for the entire system (grid plus ancilla):

$$\begin{aligned}
|\Psi\rangle &= \cos\frac{\vartheta}{2} |\psi\rangle \otimes |0\rangle + \sin\frac{\vartheta}{2} e^{i\phi} A |\psi\rangle \otimes |1\rangle \\
&= \frac{1}{\sqrt{2}} \left( \cos\frac{\vartheta}{2} \mathbb{1} + \sin\frac{\vartheta}{2} e^{i\phi} A \right) |\psi\rangle \otimes |+\rangle \\
&\quad + \frac{1}{\sqrt{2}} \left( \cos\frac{\vartheta}{2} \mathbb{1} - \sin\frac{\vartheta}{2} e^{i\phi} A \right) |\psi\rangle \otimes |-\rangle
\end{aligned} \tag{8}$$

Then, it can be shown that measuring the expectation value of the Pauli string $B \times X_a$, in which $a$ is the ancilla qubit, results in:

$$\begin{aligned}
\langle \Psi | B(t) X_a(t) | \Psi \rangle &= \sin(\vartheta)\cos(\phi)\text{Re}[\langle \psi | B(t) A(t_0) | \psi \rangle] \\
&\quad - \sin(\vartheta)\sin(\phi)\text{Im}[\langle \psi | B(t) A(t_0) | \psi \rangle]
\end{aligned} \tag{9}$$

By choosing the initial state of the ancilla to be $|\eta(\frac{\pi}{2}, 0)\rangle$, we get $\langle \psi | B(t) X_a(t) | \psi \rangle = \text{Re}[\langle \psi | B(t) A(t_0) | \psi \rangle]$. By choosing instead the ancilla in the state $|\eta(\frac{\pi}{2}, \frac{-\pi}{2})\rangle$, we get $\langle \psi | B(t) X_a(t) | \psi \rangle = \text{Im}[\langle \psi | B(t) A(t_0) | \psi \rangle]$. These initial states of the ancilla correspond to the usual version of the Hadamard test in which a Hadamard gate applied to the ancilla measures the real part of the controlled unitary and a Hadamard gate times the $S$-adjoint gate measures the imaginary part.

**Symmetry.** For the measurement of $\langle Z_l(t) Z_l(0) \rangle$ on qubit $l$, we take advantage of the symmetry of the initial state. For this correlator, separate circuits must be run for each qubit. Thus, it is a much more demanding experiment than the measurements of $\langle Z_l \rangle$, $\langle A_v \rangle$ or $\langle B_p \rangle$. To expedite the acquisition of the data shown in Fig. 4c, we only perform the measurement on 10 out of 17 physical qubits and use the vertical mirror symmetry plane to assign the values of the other seven qubits. This is reasonable because the initial state and dynamics respect this mirror symmetry and all qubits are used in the time evolution circuits for each qubit measured. Symmetrization is not used in any other figure.

**Variational quantum circuit**

In this section, we discuss the variational circuit used in the main text. In recent years, there have been other theoretical proposals based on variational methods for realizing the ground state of LGTs with shallow quantum circuits[69–71]. Here we show that our protocol is equivalent to a mean-field ansatz for the dual Ising model, for which the expectation values of the local terms of the Hamiltonian can be evaluated analytically as a function of the rotation angle $\theta$ (ref. 45). Thus, the energy optimization of the variational circuit reduces to finding the minimum of a simple polynomial of trigonometric functions, which can be solved efficiently for arbitrary system sizes. The variational state, proposed in refs. 45,46, is of the form:

$$|\psi\rangle = \prod_p \left( \cos\left(\frac{\theta}{2}\right) \mathbb{1} + \sin\left(\frac{\theta}{2}\right) B_p \right) |0\rangle, \tag{10}$$

in which the rotation angle $\theta$ is the only variational parameter. The idea behind the ansatz is to create a weight-adjustable loop gas: starting from the initial state $|0\rangle^{\otimes N}$, applying $B_p$ flips all spins around this plaquette and creates a closed loop of spins in the $|1\rangle$ state. Applying the operator $\cos\left(\frac{\theta}{2}\right) \mathbb{1} + \sin\left(\frac{\theta}{2}\right) B_p$ on a plaquette creates a weighted superposition of a closed loop and no loop around that plaquette. In the special case in which both possibilities have the same weight, that is, $\theta = \frac{\pi}{2}$, the circuit prepares the toric code ground state, which is an equal-weight superposition of all closed-loop configurations. Decreasing the angle to tune away from the toric code gives less weight to the configurations with flipped plaquettes—in particular, configurations with large loops are now exponentially suppressed, scaling with the area of the loop (that is, the number of enclosed plaquettes). Such a suppression of large loops is what we might expect when adding an onsite field term such as $-h_t \sum_i Z_i$ with $h_t > 0$ to the toric code Hamiltonian, because now long loops of flipped spins have an energy cost. However, such a suppression should scale with the perimeter of the loop, not with its area. This overpenalization of long loops leads to the fact that the ansatz cannot support topological order in the thermodynamic limit when we tune the angle away from the toric code fixed point at $\theta = \frac{\pi}{2}$. We will see this explicitly in the next section, by mapping the variational circuit to a mean-field ansatz for the dual transverse-field Ising model, in which tuning the angle away from $\theta = \frac{\pi}{2}$ means explicitly breaking the symmetry of the Ising model and thus tuning from the symmetric into the symmetry-broken phase.

**Variational circuit as a mean-field ansatz of an Ising model.** In the main text, we considered a circuit using ancilla qubits to prepare the variational ansatz because the use of ancillas then reduced the depth of the circuit needed for the time evolution. Here we are only interested in the action of the circuit on the physical system, so we can ignore the ancillas. An alternative (but equivalent) circuit for constructing the state using only the physical qubits is given in Extended Data Fig. 5a,b (refs. 42,46). Extended Data Fig. 5a shows the repeating circuit element that is applied to every plaquette. It consists of two parts: first, on the top qubit of each plaquette, a $y$-rotation gate $R_y(\theta) = \exp(-i\theta Y/2)$ is applied. Then, three C–NOT gates are applied, in which the top qubit acts as the control qubit and the remaining qubits are the target qubits. Extended Data Fig. 5b shows one possible choice of the order of applying the repeating circuit element: the element is applied to each plaquette sequentially, starting from the bottom right, traversing each row right to left, before moving up to the next row. Such an ordering ensures that the rotation gate always acts on a qubit in the $|0\rangle$ state and, thus, the circuit effectively implements the operator $\cos\left(\frac{\theta}{2}\right)\mathbb{1} + \sin\left(\frac{\theta}{2}\right)B_p$ on each plaquette, which yields the state in equation (10). Within a row, the circuit elements commute and we could choose different orderings. In fact, the presented ordering is not optimal but will simplify the calculations in the following. An optimal ordering is given in ref. 42.

To better understand the state prepared by this circuit, we can transform the Hamiltonian in the main text with only the C–NOT gates of the circuit and consider the resulting transformed Hamiltonian. The expectation value of a single Pauli-X operator evaluated in the variational state is always zero, which means that the state is completely insensitive to adding an X-field and it is enough to consider the toric code Hamiltonian with Z-field only. This is because the variational state is a superposition of closed loops only and applying a single Pauli-X operator creates open-ended loops in each state of the superposition. A schematic of the Hamiltonian is shown in Extended Data Fig. 5c. There we show a 4 × 4 lattice of vertex operators. The terms in the Hamiltonian corresponding to the vertex operators are shown in orange, the plaquette terms are shown in blue and the onsite Z-field is coloured in green. Conjugating this Hamiltonian by only the C–NOT gates of the circuit, as we will show below, leads to the Hamiltonian in Extended Data Fig. 5d. There (in the bulk) the vertex terms have shrunk to two-site Ising interactions, the plaquette terms have shrunk to onsite Pauli-X operators and the onsite Z-fields have extended to two-site or three-site Pauli-Z terms. On all sites for which there are no Pauli-X operators, the transformed Hamiltonian commutes with a single-site Pauli-Z operator, so the eigenstates of the Hamiltonian are labelled by product states of $|0\rangle$ or $|1\rangle$ on those sites. The lowest-energy states are given by $|0\rangle$ on all of those sites, as the field term in the Hamiltonian in the main text has a negative prefactor. Note that this is precisely the state of those qubits in the variational circuit before applying the C–NOT gates. Focusing on this subspace, in which all qubits except the top qubit on each plaquette are in the $|0\rangle$ state, we get the Hamiltonian in Extended Data Fig. 5e, which now only lives on the sites on which the $y$-rotation gates act in the variational circuit. We can see that, on those sites, the Hamiltonian is a two-dimensional transverse-field Ising model. The remaining variational circuit, that is, the variational circuit without the C–NOT gates, simply describes a product state with spins rotated in the $x$–$z$ plane. Such a state is a mean-field ansatz for the Ising model. In particular, if the rotation angle is $\theta = \frac{\pi}{2}$, all spins are aligned in the $x$ direction, which corresponds to the ground state of the transverse-field Ising model in the limit at which the strength of the Ising interaction is taken to zero. In that case, the ground state is symmetric under the global spin-flip symmetry of the Ising model, that is, under the simultaneous application of a Pauli-X gate on every site. However, if we tune the angle away from $\theta = \frac{\pi}{2}$, the spins are no longer oriented along the $x$ direction and the state is no longer symmetric under spin flips—there is a mean-field phase transition from the symmetric phase into the symmetry-broken phase. (Technically, the single-site Pauli-Z operators at the boundary explicitly break the global spin-flip symmetry of the Ising model in this case; however, in the thermodynamic limit, their contribution to the ground-state energy becomes negligible compared with the bulk and a transition occurs. The only effect of these boundary terms will then be to always favour the same direction of the symmetry-breaking, namely, towards the all $|0\rangle$ state). In the original model of the toric code in a field, this phase transition corresponds to a (mean-field) transition from the topological toric code ground state to the trivial paramagnet. More generally, there is a known duality transformation between the toric code with a Z-field and the two-dimensional transverse-field Ising model on the dual lattice, which relates the topological phase of the toric code to the symmetric phase of the Ising model, and the trivial phase of the toric code with a large Z-field to the symmetry-broken phase of the Ising model[1,4].

To show that the toric code Hamiltonian with a Z-field indeed transforms into an Ising model under the action of the C–NOT gates of the circuit, we first consider the action of the C–NOT gates on a single plaquette, before we then transform the full Hamiltonian plaquette by plaquette. Extended Data Fig. 6a shows all relevant operator transformations on a single plaquette. The transformation of the vertex operators on a plaquette is shown in the left column of the table. Generally, the vertex operator also includes Pauli-Z operators on neighbouring plaquettes; however, they are unaffected by the C–NOT gates applied to the selected plaquette. The right column of the table in the figure shows the transformation of the onsite Z-field. The last line at the bottom of the table shows the transformation of the plaquette operator under the C–NOT gates. Equipped with these transformation rules, we can transform the full Hamiltonian. Note that the first set of C–NOT gates that acts on the Hamiltonian is the last one that is applied to the circuit. So, to transform the Hamiltonian plaquette by plaquette, we need to proceed in the opposite order of how we constructed the circuit in Extended Data Fig. 5b. The process of the transformation is graphically depicted in Extended Data Fig. 6b. At each step, the plaquette highlighted in yellow is transformed next and each plaquette is transformed according to the rules in Extended Data Fig. 6a. The first row in Extended Data Fig. 6b shows the transformation of each plaquette in the top row of the lattice individually. In the second row, because the C–NOT gates applied on plaquettes in the same row commute, we transform a whole row in one step. The final result is the last diagram in the bottom-right corner, which is the same as the Hamiltonian in Extended Data Fig. 5d.

**Classically optimizing the variational circuit.** The mapping of the variational circuit to the two-dimensional Ising model also helps us to optimize the variational parameter $\theta$ in the ansatz, as the expectation values of all terms in the Hamiltonian can be evaluated analytically as a function of $\theta$. Minimizing the energy to find the optimal $\theta$ then reduces to finding the minimum of a simple polynomial of trigonometric functions of $\theta$.

As computed in the previous section, the vertex operators map to Ising-type interactions that only have support on sites at which the variational ansatz before the C-NOT layers is in the state $|0\rangle$—see also Extended Data Fig. 6b. Thus, the expectation values of all vertex operators in the variational state is:

$$\langle A_s \rangle = 1. \tag{11}$$

The plaquette operators map to single-site Pauli-X operators, with support on those sites at which the variational circuit has the $y$-rotation gates before the C-NOT layers. Thus, for the expectation values of the plaquette operators, we have:

$$\langle B_p \rangle = \langle 0|e^{i\theta Y/2} X e^{-i\theta Y/2}|0\rangle = \sin(\theta). \tag{12}$$

There are two different cases for the Z-field in the original Hamiltonian; in the bulk, it maps to an Ising interaction between two qubits that are acted on with a $y$-rotation gate and at the boundary, it maps to a single-site Pauli-Z operator on a qubit with a $y$ rotation. Thus, we have:

$$\langle Z_{\text{bulk}} \rangle = (\langle 0|e^{i\theta Y/2} \otimes \langle 0|e^{i\theta Y/2}) (Z \otimes Z) (e^{-i\theta Y/2}|0\rangle \otimes e^{-i\theta Y/2}|0\rangle) \\ = \cos^2(\theta) \tag{13}$$

and:

$$\langle Z_{\text{boundary}} \rangle = \langle 0|e^{i\theta Y/2} Z e^{-i\theta Y/2}|0\rangle = \cos(\theta). \tag{14}$$

As discussed in the previous section, the expectation value of single-site Pauli-X operators in the variational circuit is zero because the state is a superposition of closed loops only.

The energy of the system is then given by the sum of all terms, which for a lattice of $L_x \times L_y$ vertex operators is:

$$E = -J_E L_x L_y \\ \quad -J_M (L_x - 1)(L_y - 1)\sin(\theta) \\ \quad -h_E ((L_x - 2)(L_y - 1) + (L_x - 1)(L_y - 2))\cos^2(\theta) \\ \quad -h_E 2(L_x - 1 + L_y - 1)\cos(\theta). \tag{15}$$

This equation can be minimized efficiently numerically for arbitrary system sizes. For this, we use the default settings of the function scipy.optimize.minimize_scalar from the Python library SciPy[72], which implements Brent's algorithm[73].

In the limit $L_x, L_y \to \infty$, we can minimize the energy density $\mathcal{E} = E/(L_x L_y)$ analytically, as was similarly done in ref. 45. We only keep terms in the energy proportional to $L_x L_y$ and obtain:

$$\mathcal{E} = -J_E - J_M \sin(\theta) - 2h_E \cos^2(\theta). \tag{16}$$

Taking the derivative with respect to $\theta$, we have:

$$\frac{d\mathcal{E}}{d\theta} = -J_M \cos(\theta) + 4h_E \cos(\theta)\sin(\theta) = 0. \tag{17}$$

This equation has two solutions for $0 \leq \theta \leq \frac{\pi}{2}$. The first is given by $\cos(\theta) = 0$ or $\theta = \frac{\pi}{2}$, which corresponds to the toric code ground state, and the second is given by $\sin(\theta) = \frac{J_M}{4h_E}$ or $\theta = \arcsin\left(\frac{J_M}{4h_E}\right)$. Now, we only need to check in which regime which of the two solutions is energetically favourable. For $\theta = \frac{\pi}{2}$, we find:

$$\mathcal{E} = -J_E - J_M \tag{18}$$

and for $\theta = \arcsin\left(\frac{J_M}{4h_E}\right)$, we find:

$$\mathcal{E} = -J_E - J_M \frac{J_M}{4h_E} - 2h_E \left(1 - \left(\frac{J_M}{4h_E}\right)^2\right) \\ = -J_E - \frac{J_M^2}{8h_E} - 2h_E. \tag{19}$$

Comparing the two energies, we find that they are equal when:

$$-J_E - J_M = -J_E - \frac{J_M^2}{8h_E} - 2h_E \\ \Rightarrow 0 = \frac{J_M^2}{16h_E^2} - \frac{J_M}{2h_E} + 1 = \left(\frac{J_M}{4h_E} - 1\right)^2 \tag{20} \\ \Rightarrow h_E = \frac{J_M}{4}.$$

Thus, for $h_E \leq J_M/4$, the variational state with the lowest energy is given by $\theta = \frac{\pi}{2}$, that is, the toric code ground state, whereas for $h_E > J_M/4$, the optimized parameter is given by $\theta = \arcsin\left(\frac{J_M}{4h_E}\right)$. This is the functional form of the grey line shown in Fig. 2b.

### Further experiments

**Absolute initial state energy.** After carrying out an uncorrelated readout error mitigation procedure, the corrected values for each term in the Hamiltonian are extracted and plotted in Fig. 2d. Averaging equivalent vertices, plaquettes and links allows for a detailed understanding of how each term in $\mathcal{H}$ depends on $h_E$ (Extended Data Fig. 7).

**Heatmaps for the $|\psi_+\rangle$ state.** The ability to prepare the superposition states presented in Fig. 3c–e are a distinct advantage of digital quantum simulation. The procedure to create this superposition excitation only adds five C-NOT gates and uses a single ancilla qubit, $Q_b$ (Extended Data Fig. 8a). By first preparing a superposition between the physical system and the ancilla, the measurement of $Q_b$ in the Z basis projects the state onto either the $|\psi_+\rangle$ state or the $|\psi_-\rangle$ state. This measurement of $Q_b$ can be carried out concurrently with the measurements of all other physical and ancilla qubits.

The measurements of $\langle A_v \rangle$ across the grid show similar results for the $|\psi_+\rangle$ and $|\psi_-\rangle$ states (Extended Data Fig. 8b,c). At time $t = 0$, electric excitations are equally measured on the four sites touched by the initial excitation, with $\langle A_v \rangle \approx 0$ on each of these sites, consistent with the superposition. As time evolves, the excitations spread more for the deconfined case, $h_E = 0$, compared with the confining $h_E = 2.0$ measurements. The conditional location of charges show qualitatively different dynamics in the deconfined phase between the two initial states, which can be attributed to the different quantum interference, as discussed in the main text (Extended Data Fig. 8d). The confined conditional probabilities are similar between the two initial states.

Comparing the excitation separation measured on the device to numerical simulation shows qualitative but not quantitative agreement (Extended Data Fig. 8e). This is probably because of the cancellation of local errors, leading to worse rescaling using the global depolarizing model, as discussed for the distance 2 state in Supplementary Information Section III.B.

**Further $\langle Z(t)Z(0)\rangle$ data.** Although $\mathcal{S}_{ZZ}(t)$ and $\text{Re}[\langle Z(t)Z(0)\rangle]$ show the distinct behaviours of the string dynamics with the onset of confinement (Fig. 4), we may be interested in the behaviour of other closely related quantities to describe the string dynamics. Thus, in this section, we show experimental results and numerical simulations of $\langle Z(0)\rangle$, $\text{Im}[\langle Z(t)Z(0)\rangle]$ and $\langle Z(t)\rangle$.

Although $\langle Z(0)\rangle$ can be read from Fig. 4d as the value $\mathcal{S}_{ZZ}(0)$ (because $\text{Re}[\langle Z(0)Z(0)\rangle] = 1$), we explicitly plot those values in Extended Data Fig. 9a. In practice, these values are extracted from the same experiment that yields $\mathrm{Re}[\langle Z(t)Z(0)\rangle]$ by measuring $\langle Z_a\rangle$ for a being the ancilla qubit (standard Hadamard test).

To complement the $\mathrm{Re}[\langle Z(t)Z(0)\rangle]$ shown in Fig. 4, we present the $\mathrm{Im}[\langle Z(t)Z(0)\rangle]$ (Extended Data Fig. 9b). Oscillations are very apparent for $h_E \in \{0, 0.1, 0.2\}$ in the deconfined phase. These oscillations can be understood in the toric code limit ($\lambda$, $h_E = 0$), in which $Z$ couples the ground state to an excited state with $+4J$ higher energy. This leads to $\langle Z(t)Z(0)\rangle = \mathrm{e}^{-4iJt}$ for a bulk qubit. However, for qubits on the boundary, such as $Q_1$ and $Q_2$, one of the magnetic excitations is outside the system and do not contribute to the energy difference, therefore $\langle Z(t)Z(0)\rangle_{\mathrm{edge}} = \mathrm{e}^{-2iJt}$. The imaginary part of $\langle Z(t)Z(0)\rangle$ is suppressed as $h_E$ is increased to $h_E = 1.4$. This is also expected, because as $h_E \to \infty$, the ground state becomes an eigenstate of $Z$. Furthermore, no strong distinction between $Q_1$ and $Q_2$ is observed across the measured range of $h_E$. After rescaling assuming a global depolarizing model, good agreement is observed between experiment and numerical simulations.

Motivated by the fact that the $\mathrm{Im}[\langle Z(t)Z(0)\rangle]$ is suppressed for large $h_E$, we also measured $\langle Z(t)\rangle$ to investigate whether this observable also reveals string dynamics, because the quantum state should approach an eigenstate of $Z$ in the large $h_E$ limit. In this picture, $\langle Z_l(t)\rangle = -1/+1$ corresponds to the presence/absence of the Wilson string on qubit $l$ at time $t$. Our results show behaviour qualitatively similar to $\mathcal{S}_{ZZ}(t)$ (Extended Data Fig. 9c). We see that, for $Q_1$, the value of $\langle Z(t)\rangle$ quickly moves away from the theoretically stationary evolution (grey region) for all values of $h_E$, which is consistent with the string always moving away from its initial configuration regardless of the level of confinement. However, for $Q_2$, in the most confining case, when $h_E = 1.4$, the dynamics of $\langle Z(t)\rangle$ are indistinguishable from the stationary evolution, which corroborates our interpretation that the string is not able to move to the bottom qubits for large $h_E$. After rescaling assuming a global depolarizing model, good agreement is observed between experiment and numerical simulations.

**Temporal mapping of vacuum fluctuations and string breaking.** Having shown that, in the strongly confining phase, when $h_E = 1.4$, a string with an initial bump on the top is not able to move to qubits on the bottom of the grid on the experimentally accessible timescales owing to the small matrix elements, we presented signatures of string breaking by comparing the probabilities of finding a vertex excitation at sites at the top and at the bottom of the grid (Figs. 4 and 5). In Fig. 5, we present data after an evolution time of $t = 2.7$. Possible contributions to this late-time charge density are, for example: (1) residual energy owing to the imperfect approximation of the WALA initial state to the true ground state; (2) the disagreement between $A_v$ and the dressed particle operators for the full Hamiltonian; and (3) device decoherence.

Starting from the WALA initial state and evolving under a Hamiltonian with $\lambda = 0$, we expect no electric excitations to appear, because $A_v$ commutes with the Trotterized Hamiltonian. However, in the experiment, we see excitations developing, with the highest density on bulk sites (Extended Data Fig. 10a). This is natural for an experiment on a NISQ processor, as the noise will push the system towards the maximally mixed state, in which $\langle A_v \rangle = 0$ for all vertices. The qubits in the bulk take part in the most entangling gates and thus the effects of decoherence can be expected to be strongest for bulk sites. In this case of $\lambda = 0$, we observe equivalent results regardless of using an initial state with or without the string excitation, up to experimental errors (Extended Data Fig. 10b). When $\lambda$ is increased to 0.25 and 0.50, the trend of increasing electric excitations takes on a faster rate for the WALA state. This indicates that the noiseless evolution begins to create pairs of electric excitations owing to reasons (1) and (2) above. In the main text, we call these 'vacuum fluctuations', because they are pair-creation events that spawn from evolution of the approximate ground state. When we consider non-zero $\lambda$ for the string initial state, we see an increased excitation density compared with the WALA state. Indeed, the average probability of finding a vertex excitation on any site is markedly higher for the string initial state, compared with the WALA state, when $\lambda = 0.50$ (Extended Data Fig. 10c). After rescaling, assuming a global depolarizing model, we observe almost perfect agreement between our experimental data and noiseless numerical circuit simulations, indicating that this effect is not a spurious result of errors on the quantum processor.

Examining the heatmaps in Extended Data Fig. 10, we see that the extra intensity build-up is concentrated on vertices that the initial string passes through. The asymmetry between excitations on the top and on the bottom is the topic of Fig. 5c and represents strong evidence of string breaking. At the same time, the low probability of electric-charge creation in the WALA state by time $t = 2.7$ (about 2% for $\lambda = 0.25$ and about 5% for $\lambda = 0.50$ after depolarization mitigation) confirms that this ansatz is a good low-energy initial state with only a small amount of intrinsic dynamics.

## Data availability

The data that support the findings in this study are available on Zenodo[74].

## Code availability

Code is available on ReCirq[75].

# Article

**Acknowledgements** We acknowledge fruitful discussions with I. Aleiner and Y. Bahri. A.G.-S. acknowledges support from the Royal Commission for the Exhibition of 1851 and support from the UK Research and Innovation (UKRI) under the UK government's Horizon Europe funding guarantee (grant number EP/Y036069/1). B.J., M.W., F.P. and M. Knap acknowledge support from the Deutsche Forschungsgemeinschaft (DFG, German Research Foundation) under Germany's Excellence Strategy EXC-2111-390814868, TRR 360-492547816 and DFG grant nos. KN1254/1-2 and KN1254/2-1, the European Research Council (ERC) under the European Union's Horizon 2020 research and innovation programme (grant agreement nos. 851161 and 771537), the European Union (grant agreement no. 101169765), as well as the Munich Quantum Valley, which is supported by the Bavarian state government with funds from the Hightech Agenda Bayern Plus. M.W. and F.P. acknowledge the support of the DFG FOR 5522 Research Unit (project ID 499180199).

**Author contributions** A.G.-S., F.P., M. Knap and P.R. conceived the project. Experimental data collection protocols were conceived and implemented by T.A.C., with assistance from B.J., E.R., G.G. and N.E. Device calibration was done by T.A.C., E.R., G.G. and N.E. Theoretical models were developed by B.J., Y.D.L., M.W., A.G.-S., F.P. and M. Knap Numerical simulations were performed by T.A.C., B.J., E.R. and A. Szasz. The manuscript was written by T.A.C., B.J., E.R., Y.D.L., A.G.-S., F.P., M. Knap and P.R., with input from all of the other authors. Hardware and software contributions necessary to maintain the quantum processor were carried out by D.A., R.A., L.A.B., T.I.A., M.A., F.A., K.A., A.A., J.A., R.B., B. Ballard, J.C.B., A. Bengtsson, A. Bilmes, A. Bourassa, J.B., M.B., D.A.B., B. Buchea, B.B.B., T.B., B. Burkett, N.B., A.C., J. Campero, H.-S.C., Z.C., B. Chiaro, J. Claes, A.Y.C., J. Cogan, R.C., P.C., W.C., A.L.C., B. C., S. Das, S. Demura, L.D.L., A.D.P., P.D., I.D., A.D., A.E., A.M.E., M.E., C.E., V.S.F., L.F.B., E.F., A.G.F., B.F., S.G., R. Gasca, É.G., W.G., D. Gilboa, R. Gosula, A.G.D., D. Graumann, A.G., J.A.G., S.H., M.H., M.P.H., S.D.H., P.H., O.H., J.H., H.-Y.H., A.H., W.H., E.J., Z.J., C. Jones, C. Joshi, P.J., D.K., H.K., A.H.K., K.K., T. Khaire, T. Khattar, M. Khezri, S.K., P.K., B.K., A.K., F.K., J. Kreikebaum, V.K., D.L., T.L.-D., B. Langley, K.-M.L., J.L., K.L., B. Lester, L.L.G., W. Li, A.T.L., W. Livingston, A. Locharla, D.L., A. Lunt, S. Madhuk, A. Maloney, S. Mandrà, L.M., O.M., C.M., J.M., M.M., S. Meeks, A. Megrant, K.M., R. Molavi., S. Molina, S. Montazeri, R. Movassagh, C.N., M. Newman, A.N., M. Nguyen, C.-H.N., K.O., A.P., R.P., O.P., C.Q., G. Ramachandran, M.R., D.R., G. Roberts, K. Sankaragomathi, K. Satzinger, H.S., M.S., A. Shorter, N.S., V. Shvarts, V. Sivak, S. Small, W.C.S., S. Springer, G.S., J.S., A. Sztein, D.T., M.T., A.V., J.V., S.V., G.V., C.V.H., S.W., S.X.W., B.W., T.W., K.W., B.W.K.W., C.X., Z.J.Y., P.Y., B.Y., J.Y., N.Y., G.Y., A.Z., Y.Z., N. Zhu and N. Zobrist, under the supervision of P.R., S.B., J. Kelly, E.L., Y.C., V. Smelyanskiy and H.N.

**Competing interests** The authors declare no competing interests.

**Additional information**
**Supplementary information** The online version contains supplementary material available at https://doi.org/10.1038/s41586-025-08999-9.
**Correspondence and requests for materials** should be addressed to F. Pollmann, M. Knap or P. Roushan.
**Peer review information** *Nature* thanks Michele Burrello and the other, anonymous, reviewer(s) for their contribution to the peer review of this work. Peer reviewer reports are available.
**Reprints and permissions information** is available at http://www.nature.com/reprints.

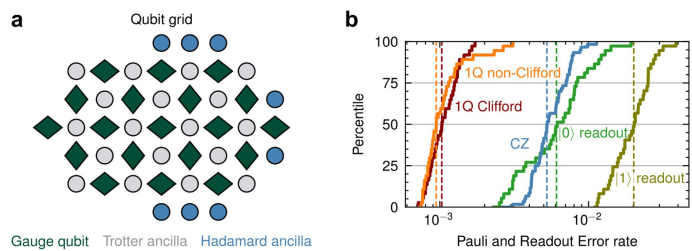

**a** Qubit grid

**b**

● Gauge qubit ● Trotter ancilla ● Hadamard ancilla

**Extended Data Fig. 1 | Qubit grid and experimental fidelities. a**, Grid of 45 qubits used in this work. Green diamonds represent physical gauge qubits, grey circles are the ancilla qubits used in Trotterized time evolution and blue circles are the ancilla qubits used in projective state preparation and Hadamard test experiments. **b**, Representative cumulative distribution functions of relevant gate and measurement errors. Single-qubit Clifford and non-Clifford Pauli errors, determined from randomized benchmarking, are shown in red and orange with median errors of 0.100% and 0.095%, respectively. To implement non-Clifford randomized benchmarking, we replace the standard depth-$n$ randomized benchmarking sequence, which consists of $n - 1$ random Clifford gates and a final Clifford gate that inverts the sequence, with $U_f X U_{\underline{n-1}} \ldots X U_0$, in which each $U_i$ is a Haar random single-qubit unitary and $U_f$ is computed to invert the whole sequence. The error rate thus obtained, which includes approximately equal contributions from the Clifford $X$ gates and the non-Clifford $U_i$ gates, is what is plotted in panel **b** as '1Q non-Clifford'. Inferred CZ Pauli errors, determined from cross-entropy benchmarking, for all pairs are shown in blue with a median error of 0.52%. $|0\rangle$ state and $|1\rangle$ state readout errors, determined from sampling random bitstrings, are shown in green and olive with median errors of 0.60% and 2.00%, respectively.

# Article

**Extended Data Fig. 2 | Suzuki–Trotter evolution circuit.** Trotterized time evolution follows initial state preparation (Fig. 2). The Trotter cycle is broken into single-qubit field terms, which act on the individual physical qubits, and plaquette terms, which involve four layers of C-NOT gates on each vertex and plaquette, single-qubit rotations on all ancilla qubits and a subsequent four layers of C-NOT gates to disentangle the ancilla qubits from the physical ones.

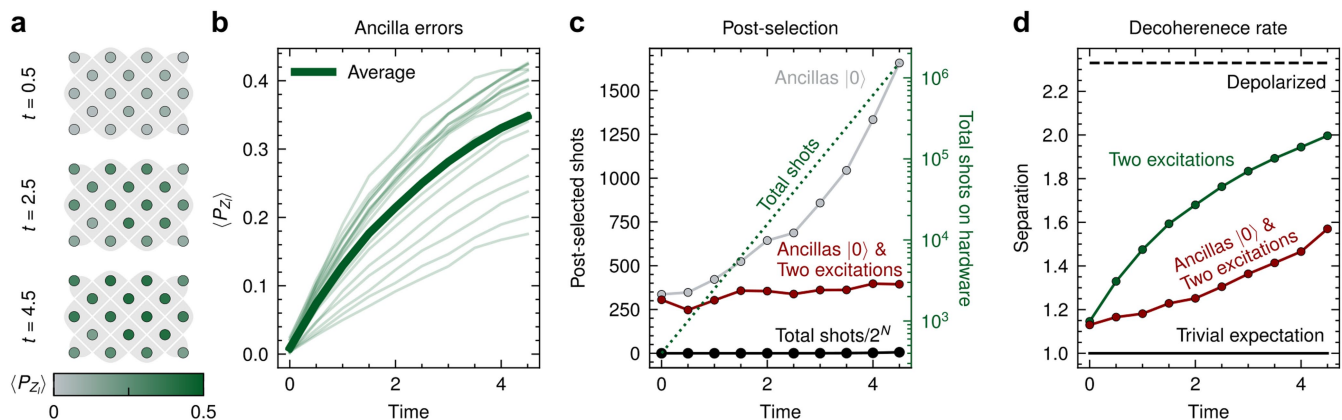

**Extended Data Fig. 3 | Post-selecting on measured ancilla state. a**, Heatmaps showing the probability of measuring the ancillas in the $|1\rangle$ state, $\langle P_{Z_l}\rangle$, at times $t \in \{0.5, 2.5, 4.5\}$ ($dt = 0.5$). A representative value of $h_L = 0.6, \lambda = 0.25$ was chosen. **b**, $\langle P_{Z_l}\rangle$ traces for all qubits (transparent traces) and their average (dark-green line). **c**, Number of total shots collected and post-selected shots as a function of evolved time. Grey points show the number of post-selected shots based on all ancillas being measured in the $|0\rangle$ state. The red points indicate the number of shots after also post-selecting on the two-excitation sector. The black points show the prediction of post-selected shots assuming that the system was in the maximally mixed state (negligible). The green dotted line (right axis) shows the total number of shots collected for each time step. **d**, Separation between two excitations, starting from the initial state shown in Fig. 3a, following evolution under the pure toric code Hamiltonian. Green markers show the separations when averaging over all bitstrings, regardless of the final state of the ancilla qubits. The red markers only average over instances when all ancilla qubits were measured in the $|0\rangle$ state. The theoretical expectation for the distance is constant at 1 (solid black line), whereas the expectation value for the maximally mixed state is 7/3 (dotted line).

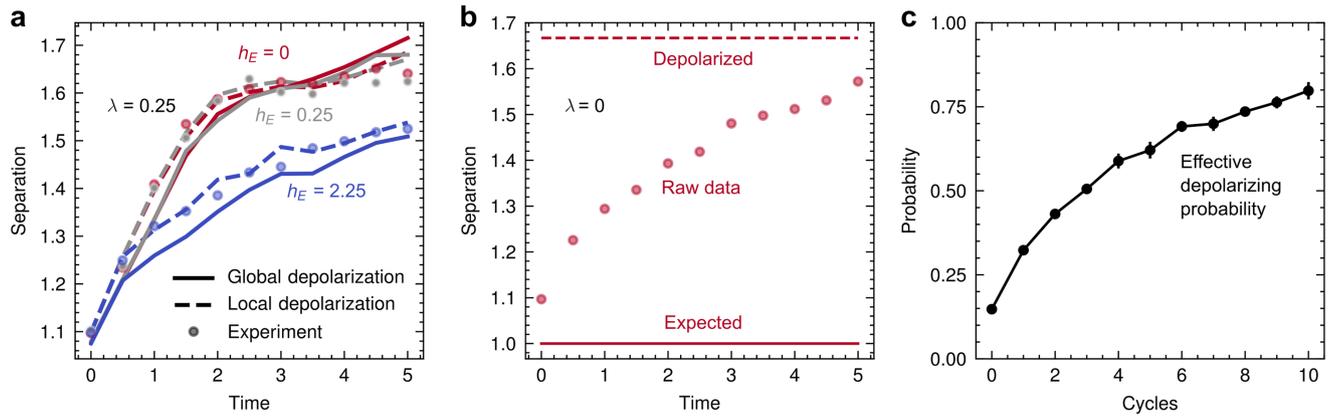

**Extended Data Fig. 4 | Local and global depolarization comparison with quantum processor data. a**, Separation between excitations, after starting with excitations a distance 1 apart on a 2 × 3 vertex lattice. Trotter evolution with $\lambda = 0.25$ and $h_E \in \{0, 0.25, 2.25\}$ are shown for device data (markers), simulations with local depolarizing noise (dashed lines) and with global depolarizing noise (solid lines). **b**, Data for evolution of the same initial state but with Trotter evolution with $\lambda = 0$, at which the vertex excitations should be stationary. The distance in the noiseless case should be 1 (solid line), whereas the expectation of distance for the maximally mixed state is 5/3 (dashed line). **c**, The extracted global depolarizing probability for the data in panel **b**.

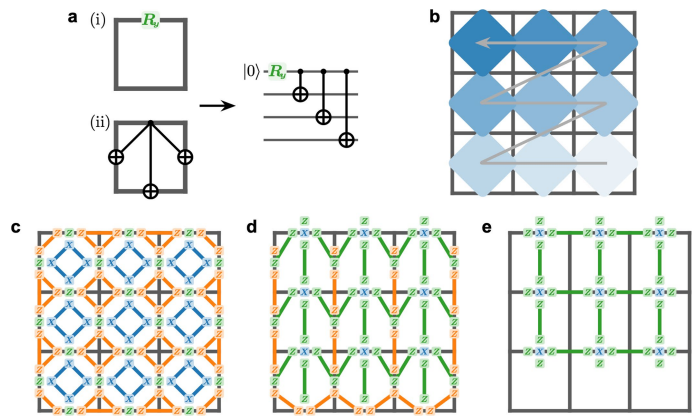

**Extended Data Fig. 5 | The variational circuit ansatz and transforming the original Hamiltonian with the C-NOT gates of the variational circuit.** This circuit is equivalent to the one in the main text but does not use ancilla qubits. **a**, The unitary applied on each plaquette consists of two parts: First, a single-qubit $y$-rotation gate $R_y(\theta) = \exp(-i\theta Y/2)$ with variational parameter $\theta$ is applied to the top qubit of the plaquette. Then, three C-NOT gates are applied, with the top qubit being the control qubit and the other qubits being the targets. **b**, The order of the plaquette unitaries is chosen such that the $y$-rotation gate always acts on a $|0\rangle$ state. Here the blue diamonds denote the gate in **a**; lighter coloured gates are applied first and darker coloured gates last. The order of plaquettes is also indicated by the grey arrow. **c**, The original Hamiltonian is drawn schematically on a lattice of $4 \times 4$ vertex operators. The orange terms connecting Pauli-Zs denote the different vertex operators of the Hamiltonian, the blue terms connecting Pauli-Xs denote the plaquette operators and the green Pauli-Zs denote the onsite Z-field. **d**, After conjugating each term in the Hamiltonian by the C-NOT layer of the circuit, we arrive at a new Hamiltonian. The orange vertex operators have been transformed to Ising terms, the blue plaquette operators have been transformed to single-site Pauli-X terms and the green Pauli-Z terms have been transformed to two-site or three-site Pauli-Z operators. On all sites at which no Pauli-X operator acts, the Hamiltonian commutes with single-site Pauli-Z operators, so on those sites, the eigenstates of the Hamiltonian are either in the $|0\rangle$ or the $|1\rangle$ state. **e**, In the subspace in which all qubits except the top qubit on each plaquette are in the $|0\rangle$ state, the transformed Hamiltonian turns into a two-dimensional transverse-field Ising model.

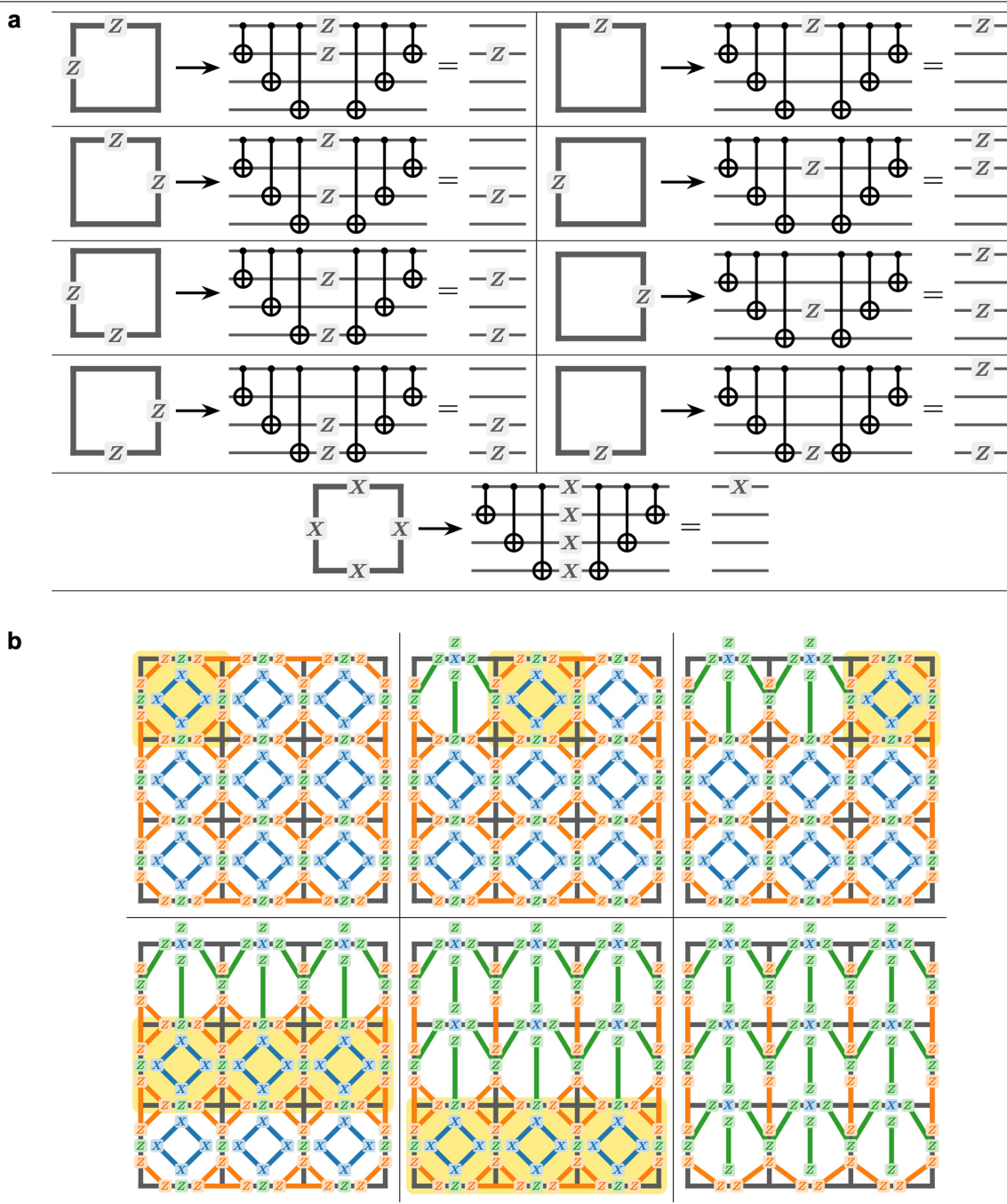

**Extended Data Fig. 6 | Operator transformations on a single plaquette and plaquette-by-plaquette transformation of the Hamiltonian. a,** The table shows the different transformations of the operators on a plaquette when transforming them with the C-NOT gates of the variational circuit acting on that plaquette. The left side shows the transformations of the different vertex operators. Note that, when a vertex operator has some remaining Z gates not supported on this specific plaquette, they are left unchanged by the C-NOT gates. The right side shows the transformations of the different onsite Pauli-Z operators. The last diagram at the bottom shows the transformation of the plaquette operators. **b,** We can use the results of the table in **a** to transform the original Hamiltonian plaquette by plaquette. At each step, the plaquette transformed next by the C-NOT gates in the variational circuit is highlighted in yellow. Note that the plaquettes of the Hamiltonian are transformed in the opposite order of how they are applied in the quantum circuit in Extended Data Fig. 5b. Because the C-NOT gates in the circuit applied to plaquettes in the same row commute, we can transform a whole row of the Hamiltonian at the same time.

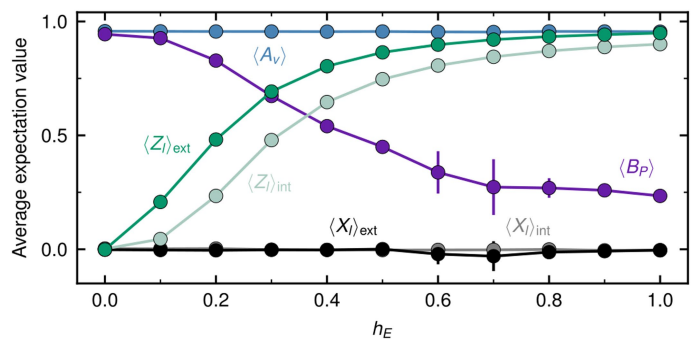

**Extended Data Fig. 7 | Average expectation values of terms in $\mathcal{H}$ for the WALA.** Expectation values of each term in the Hamiltonian, averaged over all equivalent vertices/plaquettes/links. For the expectation values of links, $\langle Z_l \rangle$ is expected to behave differently for qubits on the edge and those in the bulk (Methods). Therefore, we plot the average expectation values for these two sets of qubits separately. The expectation values are plotted for electric vertices $\langle A_v \rangle$ (blue markers), magnetic plaquettes $\langle B_P \rangle$ (purple markers), external edge links $\langle X_l \rangle_{\text{ext}}$ and $\langle Z_l \rangle_{\text{ext}}$ (dark green and black markers, respectively) and internal bulk links $\langle X_l \rangle_{\text{int}}$ and $\langle Z_l \rangle_{\text{int}}$ (light green and grey markers, respectively). Error bars correspond to the standard deviation over all vertices, plaquettes or links.

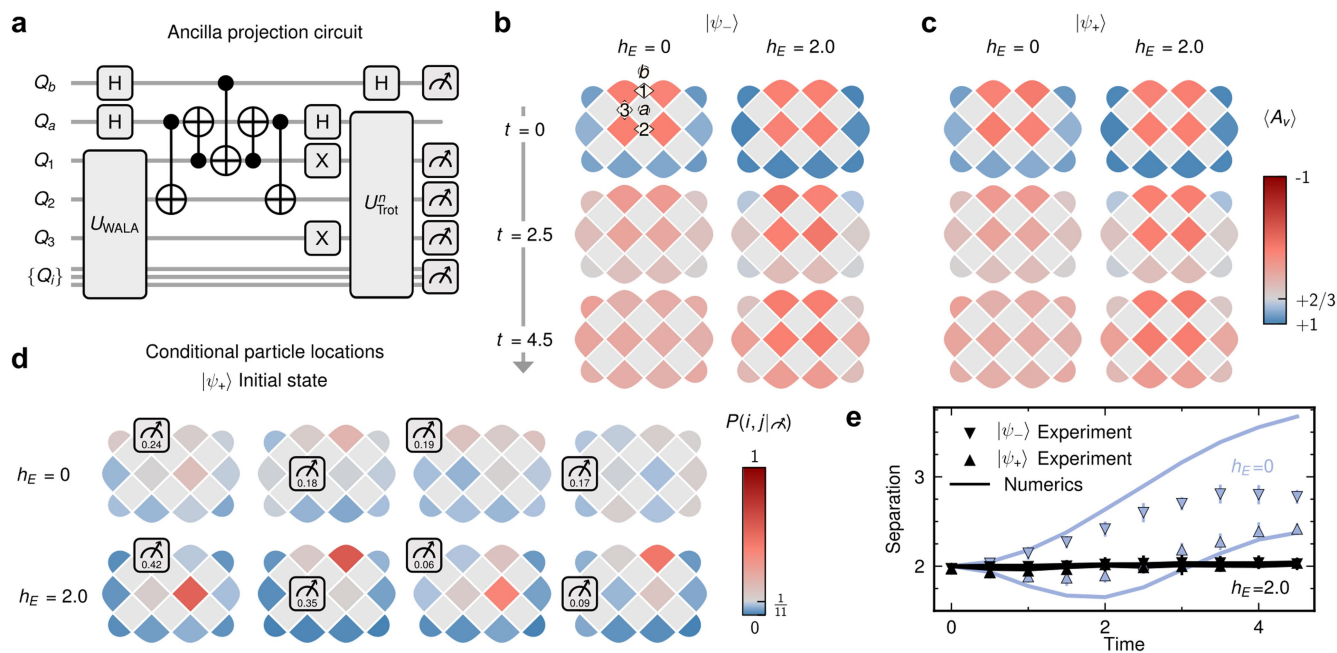

**Extended Data Fig. 8 | Average heatmaps and conditional probabilities for the superposition initial states. a**, Schematic showing the preparation of the superposition initial states. Such a circuit produces a mixed state, which can be projected on $|\psi_+\rangle$ or $|\psi_-\rangle$ depending on the measurement of the ancilla qubit $Q_b$. Qubits defined in **b**. **b**, Temporal evolution of average heatmaps of $\langle A_v \rangle$ for the $|\psi_-\rangle$ state, with $h_E \in \{0, 2.0\}$. The grey value of +2/3 on the colour bar corresponds to the average value when two electric excitations are equally distributed across the entire grid. For this figure, $\lambda = 0.25$. **c**, Temporal evolution of average heatmaps of $\langle A_v \rangle$ for the $|\psi_+\rangle$ state, with $h_E \in \{0, 2.0\}$. **d**, Conditional excitation

location probabilities for the $|\psi_+\rangle$ state, after post-selecting on the two-excitation sector, at time $t = 3.5$. The grey region of the colour bar corresponds to the average value when the excitation not conditioned on is equally distributed across the entire grid. The numbers inside the measurement boxes show the unconditioned probability of measuring an electric excitation on that site. **e**, Excitation separation for both $|\psi_\pm\rangle$ initial states and $h_E \in \{0, 2.0\}$. The markers show measured data (reproduced from Fig. 3). The lines show noiseless numerical circuit simulations.

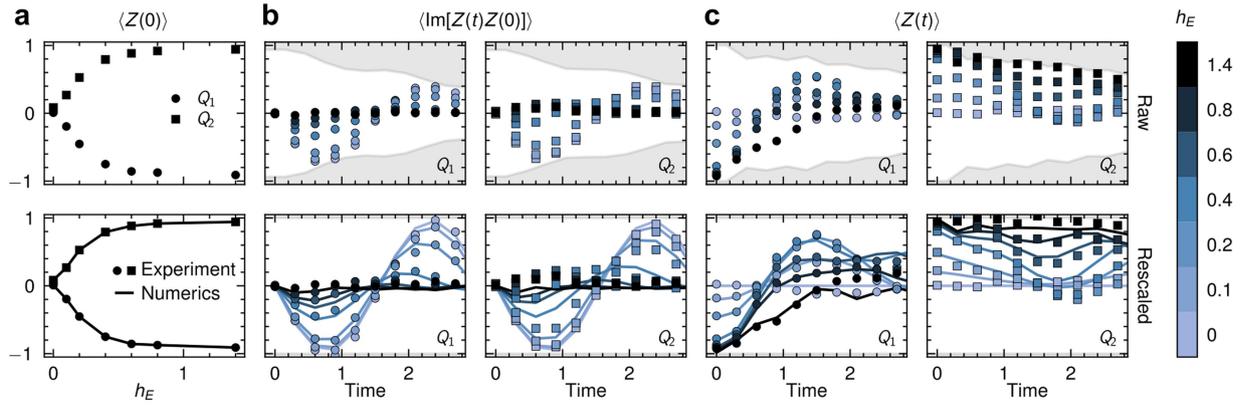

**Extended Data Fig. 9 | Measurements of $\langle Z(0)\rangle$, $\mathrm{Im}[\langle Z(t)Z(0)\rangle]$ and $\langle Z(t)\rangle$.**
**a**, Measured expectation values of $\langle Z(0)\rangle$ after preparing the WALA ground state with a string excitation as in Fig. 4a (top panel). The short-circuit depth for state preparation leads to excellent agreement with numerical simulations without any error mitigation (bottom panel). **b**, Measured expectation values of $\mathrm{Im}[\langle Z(t)Z(0)\rangle]$ for $Q_1$ and $Q_2$, defined in Fig. 4 (top panels). Data points were acquired using $dt = 0.3$ and $\lambda = 0.25$. The grey areas on these plots correspond to the region limited by decoherence and is bounded by $|\langle Z(t)Z(0)\rangle|_{\lambda=h_E=0}$.

This is then used to rescale the data to compare with noiseless numerical simulations using a global depolarizing model as described in Methods (bottom panels). **c**, Measured expectation values of $\langle Z(t)\rangle$ for $Q_1$ and $Q_2$ (top panels). The grey areas on these plots correspond to the region limited by decoherence and is bounded by $\langle Z(t)\rangle$ of the WALA initial state under evolution of the pure toric code Hamiltonian. This is then used to rescale the data to compare with noiseless numerical simulations using a global depolarizing model (bottom panels).

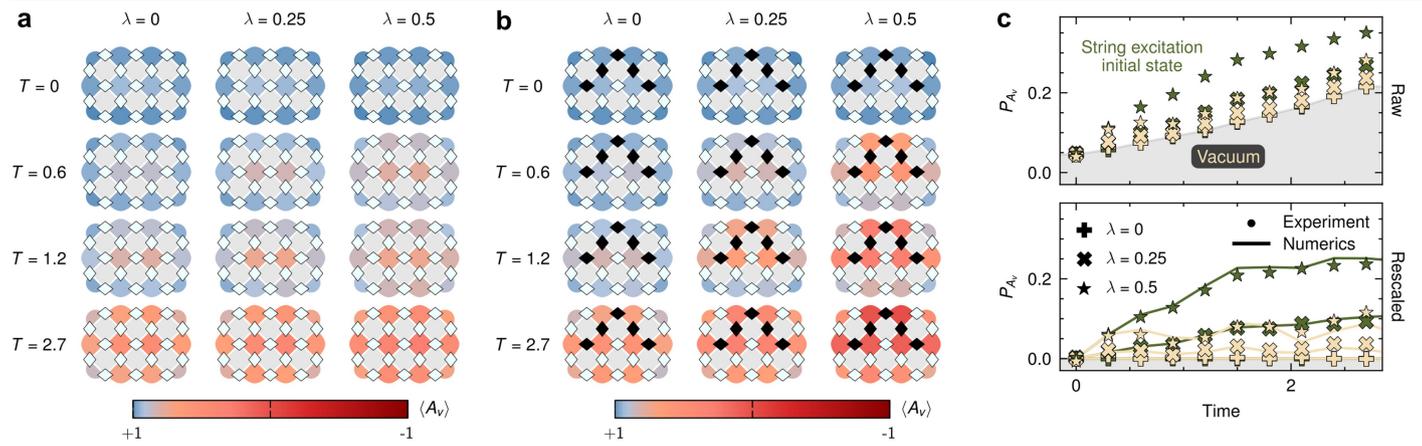

**Extended Data Fig. 10 | Build-up of vertex excitations with and without an initial string. a**, Spatiotemporal map of $\langle A_v \rangle$ for three different $\lambda \in \{0, 0.25, 0.50\}$ and constant $h_E = 1.4$, starting from the WALA initial state and time evolving. **b**, Same as panel **a** but starting with an excited initial state with a string stretched across the grid, whose initial trajectory is indicated by the black qubits. **c**, The average probability of finding a vertex excitation on any site, for each of the columns in panels **a** and **b** (both initial states). Results from evolving the WALA initial state are shown in beige and those from evolving the string initial state are shown in dark green. Markers represent experiments with $\lambda = 0$ (pluses), $\lambda = 0.25$ (crosses) and $\lambda = 0.50$ (stars). The grey region is bounded by the average of all vertices when $\lambda = 0$ having started in the initial state (same as green pluses). The bottom panel shows the global depolarization rescaled values (markers) and the numerical noiseless circuit simulations (lines).

**Supplementary information**

# Visualizing dynamics of charges and strings in (2 + 1)D lattice gauge theories

# Supplementary Information for
# Visualizing Dynamics of Charges and Strings in (2+1)D Lattice Gauge Theories

3   T. A. Cochran[1,2,‡], B. Jobst[3,4,‡], E. Rosenberg[1,‡], Y. D. Lensky[1,‡], G. Gyawali[1,5,6,‡], N. Eassa[1,7], M. Will[3,4], A. Szasz[1], D. Abanin[1],
4   R. Acharya[1], L. Aghababaie Beni[1], T. I. Andersen[1], M. Ansmann[1], F. Arute[1], K. Arya[1], A. Asfaw[1], J. Atalaya[1], R. Babbush[1], B.
5   Ballard[1], J. C. Bardin[1,8], A. Bengtsson[1], A. Bilmes[1], A. Bourassa[1], J. Bovaird[1], M. Broughton[1], D. A. Browne[1], B. Buchea[1], B.
6   B. Buckley[1], T. Burger[1], B. Burkett[1], N. Bushnell[1], A. Cabrera[1], J. Campero[1], H-S. Chang[1], Z. Chen[1], B. Chiaro[1], J. Claes[1], A.
7   Y. Cleland[1], J. Cogan[1], R. Collins[1], P. Conner[1], W. Courtney[1], A. L. Crook[1], B. Curtin[1], S. Das[1], S. Demura[1], L. De Lorenzo[1], A.
8   Di Paolo[1], P. Donohoe[1], I. Drozdov[1,9], A. Dunsworth[1], A. Eickbusch[1], A. Moshe Elbag[1], M. Elzouka[1], C. Erickson[1], V. S. Ferreira[1], L.
9   Flores Burgos[1], E. Forati[1], A. G. Fowler[1], B. Foxen[1], S. Ganjam[1], R. Gasca[1], Â. Genois[1], W. Giang[1], D. Gilboa[1], R. Gosula[1], A.
10  Grajales Dau[1], D. Graumann[1], A. Greene[1], J. A. Gross[1], S. Habegger[1], M. Hansen[1], M. P. Harrigan[1], S. D. Harrington[1], P. Heu[1], O.
11  Higgott[1], J. Hilton[1], H.-Y. Huang[1], A. Huff[1], W. Huggins[1], E. Jeffrey[1], Z. Jiang[1], C. Jones[1], C. Joshi[1], P. Juhas[1], D. Kafri[1], H. Kang[1],
12  A. H. Karamlou[1], K. Kechedzhi[1], T. Khaire[1], T. Khattar[1], M. Khezri[1], S. Kim[1], P. Klimov[1], B. Kobrin[1], A. Korotkov[1,10], F.
13  Kostritsa[1], J. Kreikebaum[1], V. Kurilovich[1], D. Landhuis[1], T. Lange-Dei[1], B. Langley[1], K.-M. Lau[1], J. Ledford[1], K. Lee[1], B. Lester[1], L.
14  Le Guevel[1], W. Li[1], A. T. Lill[1], W. Livingston[1], A. Locharla[1], D. Lundahl[1], A. Lunt[1], S. Madhuk[1], A. Maloney[1], S. Mandrâ[1], L.
15  Martin[1], O. Martin[1], C. Maxfield[1], J. McClean[1], M. McEwen[1], S. Meeks[1], A. Megrant[1], K. Miao[1], R. Molavi[1], S. Molina[1], S.
16  Montazeri[1], R. Movassagh[1], C. Neill[1], M. Newman[1], A. Nguyen[1], M. Nguyen[1], C.-H. Ni[1], K. Ottosson[1], A. Pizzuto[1], R. Potter[1], O.
17  Pritchard[1], C. Quintana[1], G. Ramachandran[1], M. Reagor[1], D. Rhodes[1], G. Roberts[1], K. Sankaragomathi[1], K. Satzinger[1], H. Schurkus[1],
18  M. Shearn[1], A. Shorter[1], N. Shutty[1], V. Shvarts[1], V. Sivak[1], V. Small[1], W. C. Smith[1], S. Springer[1], G. Sterling[1], J. Suchard[1], A.
19  Sztein[1], D. Thor[1], M. Torunbalci[1], A. Vaishnav[1], J. Vargas[1], S. Vdovichev[1], G. Vidal[1], C. Vollgraff Heidweiller[1], S. Waltman[1], S.
20  X. Wang[1], B. Ware[1], T. White[1], K. Wong[1], B. W. K. Woo[1], C. Xing[1], Z. Jamie Yao[1], P. Yeh[1], B. Ying[1], J. Yoo[1], N. Yosri[1], G. Young[1],
21  A. Zalcman[1], Y. Zhang[1], N. Zhu[1], N. Zobrist[1], S. Boixo[1], J. Kelly[1], E. Lucero[1], Y. Chen[1], V. Smelyanskiy[1], H. Neven[1], A.
22  Gammon-Smith[11,12], F. Pollmann[3,4,§], M. Knap[3,4,§], P. Roushan[1,§]

23  1 Google Research, Mountain View, CA, USA
24  2 Department of Physics, Princeton University, Princeton, NJ, USA
25  3 Technical University of Munich, TUM School of Natural Sciences, Physics Department, 85748 Garching, Germany
26  4 Munich Center for Quantum Science and Technology (MCQST), Schellingstr. 4, 80799 München, Germany
27  5 Department of Physics, Cornell University, Ithaca, NY, USA
28  6 Laboratory of Solid State and Atomic Physics, Cornell University, Ithaca, NY, USA
29  7 Department of Physics and Astronomy, Purdue University, West Lafayette, IN 47906, USA
30  8 Department of Electrical and Computer Engineering, University of Massachusetts, Amherst, MA
31  9 Department of Physics, University of Connecticut, Storrs, CT
32  10 Department of Electrical and Computer Engineering, University of California, Riverside, CA
33  11 School of Physics and Astronomy, University of Nottingham, NG7 2RD, UK
34  12 Centre for the Mathematics and Theoretical Physics of Quantum Non-Equilibrium Systems, University of Nottingham, Nottingham, NG7 2RD, UK

35  ‡ These authors contributed equally to this work.
36  § Corresponding author: frank.pollmann@tum.de
37  § Corresponding author: michael.knap@ph.tum.de
38  § Corresponding author: pedramr@google.com

# I. List of symbols

| Symbol | Description |
|---|---|
| $\mathcal{H}$ | $\mathbb{Z}_2$ lattice gauge theory Hamiltonian in (2+1)D |
| $Q_l$ | Qubit on site $l$ |
| $X_l, Y_l, Z_l$ | Spin-1/2 Pauli operators on the qubit that lives on link $l$ |
| $A_v$ | Vertex operator acting on vertex $v$, $A_v = \prod_{i \in v} Z_i$ |
| $B_p$ | Plaquette operator acting on plaquette $p$, $B_p = \prod_{i \in p} X_i$ |
| $J_E$ | Energy scale of the electric (vertex) excitations |
| $J_M$ | Energy scale of the magnetic (plaquette) excitations |
| $h_E$ | Electric field |
| $\lambda$ | Matter-gauge coupling |
| $N$ | Total number of link qubits |
| $N_{A_v}$ | Total number of electric vertex sites |
| $N_{B_p}$ | Total number of magnetic plaquette sites |
| $L_x, L_y$ | Number of vertices along $x$ and $y$ directions, respectively |
| $dt$ | Trotter step size |
| $n$ | Number of Trotter steps |
| $\theta$ | Angle of the initial rotation around the $Y$-axis of the ancilla qubit in |
|  | the Weight Adjustable Loop Ansatz (WALA) |
| $\vartheta, \phi$ | Ancilla qubit rotation angles for implementing Hadamard test |
| $h_{\mathrm{mf}}$ | Value of $h_E$ at $\lambda = 0$ corresponding to the mean-field phase transition |
| $|\psi_\pm\rangle$ | Positive and negative superpositions of electric excitations separated by Manhattan distance two |
| $U_{\mathrm{WALA}}$ | Parameterized Weight Adjustable Loop Ansatz (WALA) used to prepare the |
|  | low-energy initial state $|\psi_0(\theta)\rangle$ |
| $U_{\mathrm{Fields}}$ | Unitary corresponding the the application of the single qubit field terms of $\mathcal{H}$ |
| $U_{\mathrm{Plaquettes}}$ | Unitary corresponding the the application of the vertex and plaquette terms of $\mathcal{H}$ |
| $U_{\mathrm{Trot}}$ | Floquet unitary $\exp(-i\mathcal{H}dt)$ corresponding to a single cycle of Trotterized dynamics |
| $\mathcal{S}_{ZZ}(t)$ | String dynamics correlation function given by $\mathrm{Re}[\langle Z(t)Z(0)\rangle] \times \langle Z(0)\rangle$ |
| $\mathcal{C}(j,t)$ | Two-time string correlator given by $\langle\psi_0|(X_{Q_1}X_{Q_2}....X_{Q_j})(t)X_{Q_1}(0)|\psi_0\rangle$ |
| $P_{A_v}$ | Probability of a vertex $v$ being excited |
| $P_{Z_l}$ | Probability of measuring qubit on link $l$ in the $|1\rangle$ state |
| $p_{\mathrm{eff}}$ | Effective depolarization probability of the global depolarizing channel |
| $\mathcal{E}$ | Energy density $E/(L_xL_y)$ |

## II.  Approximation error of the WALA state

In the Methods section of the main text, we presented the method we used to optimize the angle $\theta$ of the WALA circuit for the experimental system consisting of $3 \times 4$ vertex operators, which gives the angles shown in Fig. 2b in the main text. Here, we assess the quality of the initial state for this finite system by considering the energy difference from the exact ground state and the fidelity obtained via exact diagonalization. The results are shown in Fig. S1. The left plot shows the relative energy difference, defined as $|E_{\text{exact}} - E_{\text{WALA}}|/|E_{\text{exact}}|$, and the right plot shows the infidelity, defined as $1 - |\langle \psi_{\text{exact}} | \psi_{\text{WALA}} \rangle|^2$, as the strength of the electric field $h_E$ is tuned for both $\lambda = 0$ (blue) and $\lambda = 0.25$ (orange). For $\lambda = 0$, the WALA state is just a mean-field ansatz for the dual Ising model and we observe both a very small energy difference and infidelity below $10^{-2}$ for the whole parameter regime. The WALA state does not change when turning on a small coupling $\lambda = 0.25$, since it is insensitive to an onsite $X$-field. However, while the Hamiltonian is no longer dual to an Ising model, the perturbation is weak and thus the WALA state should still be a low-energy density state. This is confirmed by the small energy difference and infidelity below $10^{-1}$ throughout the whole parameter regime shown in the figure.

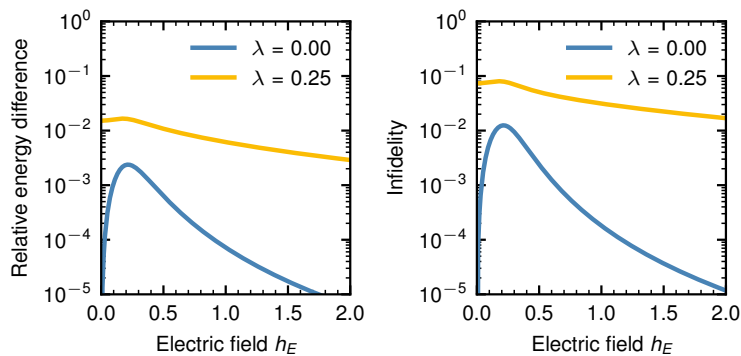

FIG. S1. **Relative energy difference and infidelity of the WALA state compared to the exact ground state for the experimental system size of 3 × 4 vertices.** The left plot shows the relative energy difference $|E_{\text{exact}} - E_{\text{WALA}}|/|E_{\text{exact}}|$ as the electric field $h_E$ is tuned both for $\lambda = 0$ (blue), where the mapping to the Ising model exists, and for $\lambda = 0.25$ (orange), for which most of the data in the main text is presented. The right plot shows the infidelity, defined as $1 - |\langle \psi_{\text{exact}} | \psi_{\text{WALA}} \rangle|^2$, against the electric field $h_E$ for both $\lambda = 0$ (blue) and $\lambda = 0.25$ (orange).

## III. Further experimental data

### A. Absolute initial state energy

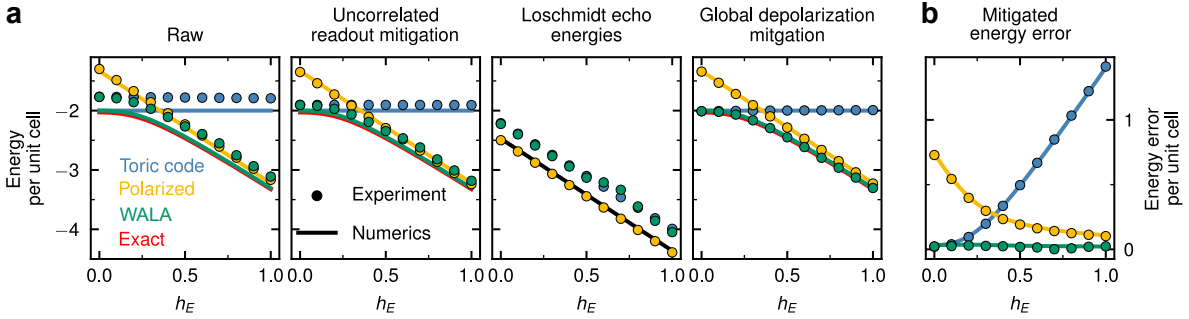

FIG. S2. **Energy of the WALA, toric code, and polarized initial states. a**, Raw energy of the initial states after post-selecting on all ancilla qubits being measured in the $|0\rangle$ state. Points correspond to experimentally measured values, whereas lines correspond to theoretical values. The red line shows the energy for the ground state found from exact diagonalization. Uncorrelated readout error mitigation results show reduced deviation from the exact values. These are the results that are presented in Fig. 2 of the main text. Performing a Loschmidt echo of the state preparation circuit and computing the resulting energy of the final state allows for the extraction of a depolarization rescaling parameter. After the Loschmidt circuit, all of the initial states should correspond to the black line in the noise-free limit. The rescaled results show almost perfect agreement with theoretical expectations. **b**, Energy error, compared to the exact diagonalization results, after the global depolarization rescaling.

The WALA initial state has been established as a suitable low-energy initial state that approximates the ground state and approaches the exact ground state in the limits of $h_E \to 0$ and $h_E \to \infty$ (Methods). To confirm this holds true after preparing the WALA circuit on the quantum processor, we prepare three initial states. The toric code initial state corresponds to the exact ground state of the toric code Hamiltonian, equivalent to the WALA initial state with $\theta = \pi/2$. The polarized initial state is the product state with all qubits in the $|0\rangle$ state. Whereas the WALA state follows the preparation described in the main text Fig. 2, with optimized initial angle determined by the procedure in the Methods. After state preparation, measurements are taken in either the $Z$ or $X$ basis and the energy is computed by directly determining the expectation value of each term in the Hamiltonian from the acquired bitstrings. Since only the WALA state depends on $h_E$, the data for the polarized and toric code initial states was collected only once. The different points correspond to the total energy calculated from the measured observables using different values of $h_E$.

The raw results of the energy per unit cell are shown in Fig. S2. The toric code and WALA results both show offsets compared to the theoretical expectation values at the level of about 0.2 $J_E$ per unit cell, whereas the polarized state shows almost perfect agreement with the expected value. The smaller deviation from expected energies in the polarized state can be attributed to lower decoherence from the trivial circuit to prepare the state (qubits are already initialized in the $|0\rangle$ state with high fidelity), compared to the five CZ layer depth to prepare the WALA or toric code states. There is also a contribution to the deviation from readout error, which will be least for the polarized state, since the $|0\rangle$ state readout error is on the order of 3–4 times smaller than the $|1\rangle$ error (Methods). Because we are interested in the ability to prepare a low energy initial state for subsequent dynamics, we report the energy values after correcting for readout errors in the main text.

To mitigate readout errors, we construct individual readout confusion matrices $\mathcal{R}_Q$ for each qubit:

$$\mathcal{R}_Q = \begin{pmatrix} 1 - \epsilon_{Q,0} & \epsilon_{Q,1} \\ \epsilon_{Q,0} & 1 - \epsilon_{Q,1} \end{pmatrix} \tag{S1}$$

where $\epsilon_{Q,0}/\epsilon_{Q,1}$ is the error of measuring the $|0\rangle/|1\rangle$ state on qubit $Q$, respectively, as determined by sampling random bitstrings. Then, to compute the readout-mitigated value of any length $n$ Pauli string, in this case an individual term in $\mathcal{H}$, we construct a confusion matrix for that multi-qubit observable, $\mathcal{R}_{\mathcal{O}}$, assuming uncorrelated readout errors:

$$\mathcal{R}_{\mathcal{O}} = \mathcal{R}_{Q1} \otimes \mathcal{R}_{Q2} \otimes \ldots \otimes \mathcal{R}_{Qn} \tag{S2}$$

such that $\mathcal{R}_{\mathcal{O}}$ is a $2^n \times 2^n$ matrix. Then, taking $\overrightarrow{P}$ to be the probability vector in the computational basis of the qubits in the support of $\mathcal{O}$, we perform the readout error mitigation:

$$\overrightarrow{P}_{\text{mitigated}} = \mathcal{R}_{\mathcal{O}}^{-1} \overrightarrow{P}_{\text{measured}} \tag{S3}$$

which can then be utilized to compute the readout-mitigated expectation value of $\mathcal{O}$.

After carrying out this uncorrelated readout error mitigation procedure, the corrected values for each term in the Hamiltonian are extracted and plotted in Fig. 2d. The discrepancy between experiment and theory is narrowed by readout error mitigation for all three initial states, but the largest effect is seen in the toric code and WALA states, consistent with the $|1\rangle$ errors playing a more significant roles in these states. The residual difference between theory and experiment is attributed to decoherence during the state preparation and represents a small error. We note a bias for the polarized state, which arises because of a small deviation from perfectly uncorrelated readout errors of the all $|0\rangle$ state. This results in the readout-mitigated values being slightly lower in energy than the noiseless simulation. Crucially, comparing the energy error of these readout-mitigated states we note that the experimental value of the WALA state is always less than or equal to the measured energy for either of the other initial states, and is also less than the smallest energy scale in $\mathcal{H}$: $\lambda = 0.25$ (main text Fig. 2c).

The readout-mitigated data are most relevant for the ground state preparation and are presented in the main text Fig. 2. However, it is also reasonable to ask how precisely we can extract the ground state energy itself using error mitigation. We do this by utilizing a Loschmidt echo technique similar to that used in [1, 2] to obtain an effective global depolarization rescaling parameter.

To start, we note that the energy we measure on the device, $E_{\text{measured}}$, is the sum of the expectation values of each term in $\mathcal{H}$. We obtain these terms by measuring all qubits in the computational basis after two circuits: (1) $U_{\text{prep}}$ gives the expectation values of $\langle Z_l \rangle$ and $\langle A_v \rangle$ for all $l, v$, and (2) $(\prod_l \text{H}_l) U_{\text{prep}}$ gives $\langle X_l \rangle$ and $\langle B_p \rangle$ for all $l, p$ ($\text{H}_l$ is a Hadamard gate on qubit $l$). Due to depolarizing noise, the expectation value for each operator, $\mathcal{O}$, will approach the value of the maximally mixed state, $\text{Tr}\{\mathcal{O}\mathbb{1}\}$, exponentially as the noise in the circuit is increased. Since each term in $\mathcal{H}$ is traceless, $\text{Tr}\{\mathcal{O}\mathbb{1}\} \to 0$ as noise is increased and thus $E_{\text{measured}} \to 0$, exponentially as well. Employing Eq. 6,

$$E_{\text{rescaled}} = \frac{E_{\text{measured}}}{1 - p_{\text{eff}}},\tag{S4}$$

for some effective depolarizing probability $p_{\text{eff}}$, which must be determined from another measurement. Our assumption is that $p_{\text{eff}} = e^{-\gamma d}$, where $\gamma$ is a constant and $d$ is the depth of the circuit. This simple model results in very good agreement between experiment and numerical simulations. Considering the Loschmidt circuit, $U_{\text{prep}}^\dagger U_{\text{prep}}$, we expect a depolarizing probability of $p_{\text{Loschmidt}} = p_{\text{eff}}^2$ since the depth doubles. Also, since $U^\dagger U |0\rangle^{\otimes N} = |0\rangle^{\otimes N}$, the expectation value for each of the $Z$-basis operators is trivial after the Loschmidt evolution: $\langle Z_l \rangle_{\text{Loschmidt}} = \langle A_v \rangle_{\text{Loschmidt}} = +1$ for all $l, v$. Similarly, applying the Loschmidt echo $U_{\text{prep}}^\dagger (\prod_l \text{H}_l)(\prod_l \text{H}_l) U_{\text{prep}}$ and measuring in the $Z$-basis yields $\langle X_l \rangle_{\text{Loschmidt}} = \langle B_p \rangle_{\text{Loschmidt}} = +1$ for all $l, p$. Therefore the sum of all the terms in $\mathcal{H}$ after the Loschmidt echo should yield $E_{\text{Loschmidt,exact}} = -(\lambda + h_E) \times N - J_E \times N_{A_v} - J_M \times N_{B_p}$. Given this exact (noiseless) value and the noisy results from our the quantum processor (Fig. S2a), Eq. 5 in the Methods sectionsstipulates that

$$p_{\text{Loschmidt}} = 1 - \frac{E_{\text{Loschmidt,measured}}}{E_{\text{Loschmidt,exact}}}.\tag{S5}$$

Finally,

$$p_{\text{eff}} = \sqrt{p_{\text{Loschmidt}}} = \sqrt{1 - \frac{E_{\text{Loschmidt,measured}}}{E_{\text{Loschmidt,exact}}}},\tag{S6}$$

which can be plugged into to Eq. S4 to yield the error-mitigated energy. The deviation from the true ground state energy shows good agreement after this rescaling procedure (Fig. S2b).

## B. Simple initial states with two charges

The choice of where two electric excitations begin impacts both their ideal dynamics and the severity of Trotter error. In the main text Fig. 3a,b, we show the dynamics after the excitations start directly next to each other with a relatively small $dt = 0.3$. The advantage of the small Trotter step is that we observe the fine dynamics of the charges' motion. Zooming in on the $h_E = 2.0$ case, the oscillations are indicative of confined charges and match the numerical circuit simulation very well (Fig. S3a). In this strongly confined regime, we expect such oscillations arising from the coherent motion of excitations in a linearly slanted potential on a lattice.

Increasing the Trotter step to $dt = 0.5$ allows for later times to be simulated, when distinctions between our chosen values of $h_E$ are clearer (Fig. S3b). Indeed, we see signatures of confinement at $h_E = 0.6$, where the separation between excitations levels off and starts to decrease before the expected distance of the maximally mixed state of 7/3. For $h_E = 0.8$, the excitations move back together even sooner. The drawback of the $dt = 0.5$ data is the non-negligible Trotter error for larger $h_E$ (Supplementary Information IV C). This Trotter error not only kills the coherent oscillations seen for $dt = 0.5$, but also causes the confined excitations to slowly drift apart from spurious hopping, as seen for $h_E = 2.0$.

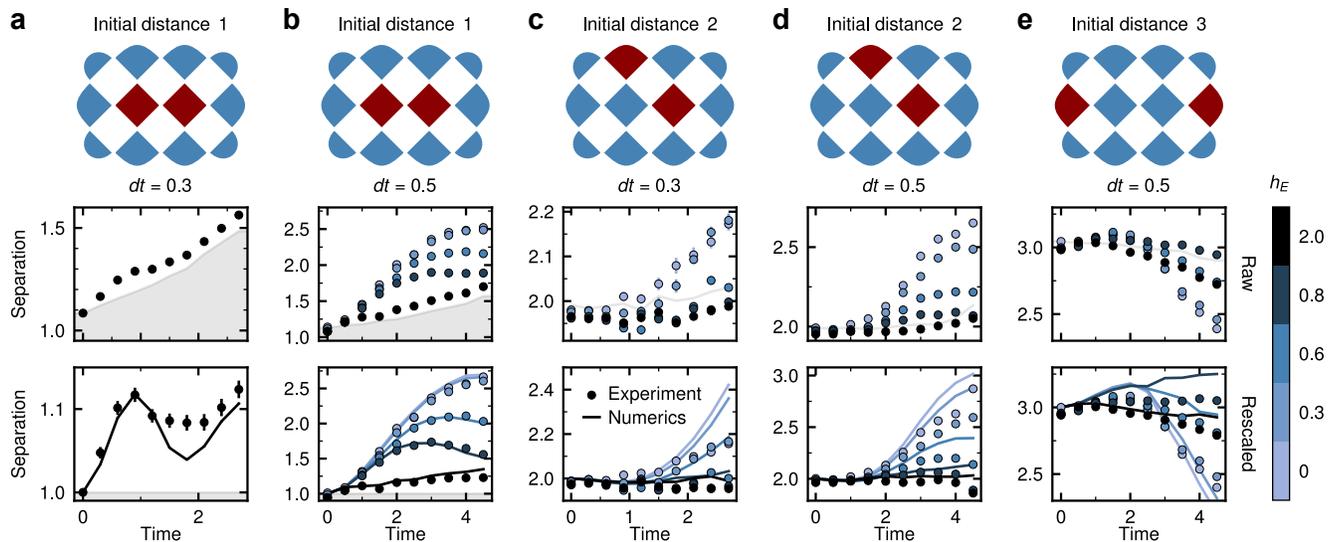

FIG. S3. **Electric excitation separation dynamics of simple initial states.** Separation between electric excitations as a function of time for $h_E \in \{0, 0.3, 0.6, 0.8, 2.0\}$. In each panel, the top row of schematics represent the initial state of two electric excitations of the WALA initial state. The middle panel shows raw data with the grey region/line being the depolarization-limited region, bounded by evolution under the pure toric code Hamiltonian. The lower panel shows the rescaled data (assuming a global depolarizing channel) and numerical simulations. Each panel varies by initial state and Trotter step size, $dt$, as follows: **a**, Initial separation 1, $dt = 0.3$. For this panel, the grey region is bounded by evolution under the Hamiltonian with $\lambda = 0$, $h_E = 2.0$, since we are only showing the dynamics for $h_E = 2.0$. This curve is also used to rescale the experimental data in the lower panel. **b**, Initial separation 1, $dt = 0.5$. **c**, Initial separation 2, $dt = 0.3$. **d**, Initial separation 2, $dt = 0.5$. **e**, Initial separation 3, $dt = 0.5$.

As mentioned in Supplementary Information IV C, by choosing an initial state with excitations two sites apart, the leading order effect of Trotter error on separation effectively cancels out. Therefore, the most confined evolution maintains a separation of 2 even for $dt = 0.5$. However, an analogous cancellation of local hardware errors can occur. Such a cancellation could be responsible for the remarkable stability of the charge separation under evolution by the pure toric code Hamiltonian (grey lines in Fig. S3c,d), where the $A_v$ operators commute with the Hamiltonian. This is in stark contrast to the experiments with initial separation of 1, where the pure toric code evolution results in a marked drift of charges towards the expectation value of the maximally mixed state of $7/3$. Therefore, we may expect that using the toric code evolution to calibrate the global depolarization rescaling may yield a poorer result for larger initial distances, which is what we observe in experiment.

We also explore dynamics after starting excitations on either edge, a distance 3 apart (Fig. S3e). In this case the confined excitations remain close to their initial separation of 3 due to the conservation of energy on the lattice.

## C. Dynamics of a single mobile excitation

To gain additional insight into confinement, we can utilize our precise local Hamiltonian control to visualize the dynamics of a single excitation. Such an approach allows us to disentangle the motion of the other excitation to narrow down the possible number of configurations the system can take on. We explore this approach by measuring both vertex and string excitations.

### 1. Measurement of a single vertex excitation

To start, let us consider a set of physical qubits denoted $\mathcal{Q}$. We consider an electric excitation pair created at an edge vertex $A_{\text{edge}}$ (Fig. S4a). We first add an extra qubit, $Q_0$, to promote the two-qubit term $A_{\text{edge}} = \prod Z_l$ to a three-qubit term $\tilde{A}_{\text{edge}} = \prod Z_l$, while adding an additional electric site with Hamiltonian term $A_{\text{pinned}} = Z_{Q_0}$. This extra qubit does not effect the ground state circuit, which only acts around magnetic plaquette sites. Then, applying an $X$ gate to $Q_0$, we create a pair of excitations located on $A_{\text{pinned}}$ and $\tilde{A}_{\text{edge}}$. By not applying the $\lambda$, $h_E$ local field terms to $Q_0$, we pin the outer excitation on $A_{\text{pinned}}$, isolating the dynamics of the one inner excitation, originally on $\tilde{A}_{\text{edge}}$. With this setup, the quantum state remains a product of $|\phi_{Q_0}\rangle \otimes |\psi_Q\rangle$, with $|\phi_{Q_0}\rangle = |1\rangle$. When considering the effect of measurement in the $Z$ basis, the $A_{\text{pinned}}$ term becomes a constant with $\langle A_{\text{pinned}} \rangle = -1$ and $\tilde{A}_{\text{edge}} = -A_{\text{edge}}$. The result is an electric flux line entering the system from the boundary at $Q_0$.

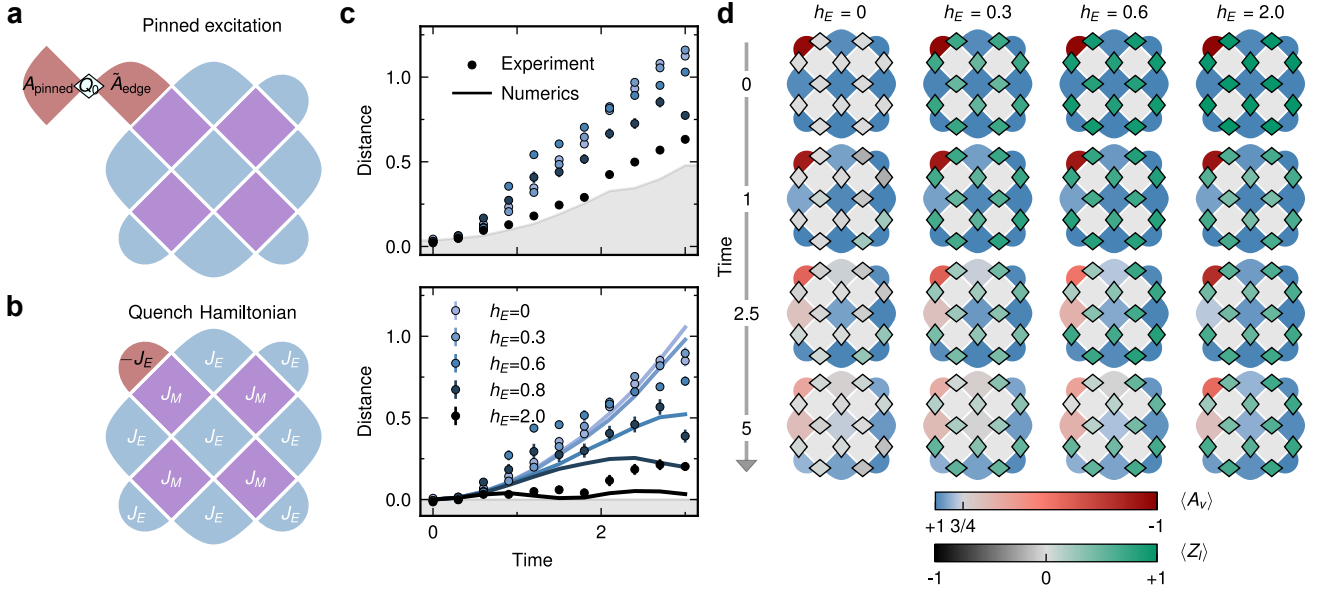

FIG. S4. **Dynamics of a single mobile excitation. a,** Standard rectangular grid configuration with an extra qubit, $Q_0$ incorporated to allow for a vertex site $A_{\text{pinned}}$ adjacent to $\tilde{A}_{\text{edge}}$. **b,** Schematic for the quench Hamiltonian equivalent to the initial state in **a**. The sign of $J_E$ is flipped just on the $A_{\text{edge}}$ vertex. **c,** Distance from $A_{\text{edge}}$ that a vertex excitation is measured after post-selecting for one interior excitation, acquired with $\lambda = 0.25$ and $dt = 0.3$. The top panel shows unmitigated results with the grey region bounded by the measured distance under the evolution of the pure toric code Hamiltonian, where all $A_v$ commute with the Hamiltonian. The bottom panel shows the global depolarization rescaled results along with numerical simulations. **d,** Spatio-temporal heatmaps of excitations for several values of $h_E \in \{0, 0.3, 0.6, 2.0\}$. The expectation value of the charge vertices, $\langle A_v \rangle$, are shown by the red/blue colorbar. The expectation value $\langle Z_l \rangle$ for each gauge qubit is shown by the green/black colorscale.

Since the system remains in a product state between $Q_0$ and $Q$, we can imagine simulating just the dynamics of $Q$ alone. To this end, we take advantage of the fact that the dynamics of the initial state shown in Fig. S4a are equivalent to a quantum quench using a Hamiltonian with $J_E \to -J_E$ on the $A_{\text{edge}}$ vertex, acting only on $Q$ (Fig. S4b). Thus we can effectuate the dynamics of a single vertex excitation without increasing the number of qubits or two-qubit gates. In our minds, we can remember the pinned excitation, which would be connected to the interior excitation by a string. Indeed the movement of the excitation in the quench protocol will be accompanied by a string leading back to its initial position (for the confining case). Interestingly, this quantum quench also removes any constraints, imposed by qubit connectivity, on introducing a single charge at any site in the bulk or on the edge. Such a scheme could be used to study interactions between excitations in the bulk, which are not initially linked by a string.

By analyzing the average distance an excitation moves away from its initial state, signatures of confinement are indeed evident (Fig. S4c). Similar to the separation between two mobile charges, we observe a clear trend in the distance the one charge travels from its original location with $h_E$. When $h_E$ is small, the excitation quickly moves away from its initial location. This indicates the charge explores the grid. However, when $h_E = 0.6$ the average distance the charge has moves evolves slower. With increasing $h_E$, the confinement to the excitation's initial position becomes stronger. In the fully confined phase with $h_E = 2.0$, we observe the charge as almost stationary, after rescaling for decoherence using a global depolarizing model. The results show good agreement between experiment and exact circuit simulation.

In Fig. S4d, we provide the full spatiotemporal mapping of the dynamics of a single excitation and the resulting *electric field* on each gauge qubit, $\langle Z_l \rangle$. For parameters that place the dynamics far outside the toric code phase, i.e. when $h_E = 2.0$ and $\lambda = 0.25$, the electric excitation does not move far from its initial position on the top left of the grid. The excitation staying at its initial site signifies confinement. This can be understood as a screening effect of the flux string that enters the system from the pinned external charge. Indeed, $\langle Z_l \rangle$ is nearly constant during our evolution time with only small deviations in the upper left corner where the flux is threaded in. Near the toric code phase, i.e. when $h_E = 0$ and $\lambda = 0.25$, the excitation shows clear indications of deconfinement, with the probability of finding the charge on the initial site quickly falling as the probability of the charge occupying nearby sites increases, indicating free diffusion. A complementary view comes from observing that $\langle Z_l \rangle \sim 0$ for all measurements, consistent with no electric flux string existing in the deconfined phase. When $h_E = 0.6$ and $\lambda = 0.25$, which we expect to be in the confined phase, we observe signatures of weak confinement. There is some diffusion to nearby sites, but the overall tendency is for the excitation to stay in the upper left corner. For this value of external field, we observe the electric flux only partially percolates into the bulk, consistent with the formation of a charge screening cloud.

## 2. Measurement of a pinned $X$-string

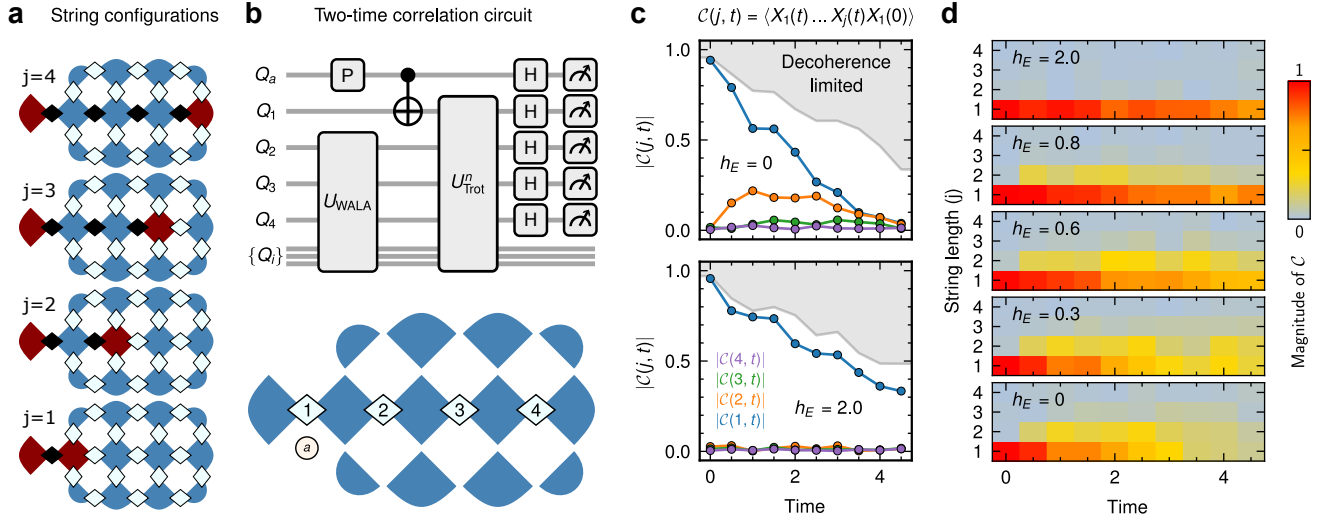

FIG. S5. **String two-time correlator. a**, Schematics of the string configurations to be considered. The string begins on the extra vertex site on the left and stretches $j$ sites to the right. **b**, Schematics for the modified Hadamard test circuit used to measure $\mathcal{C}$. Qubits $Q_a, Q_1, \ldots, Q_4$ are labeled in the schematic near the bottom of the panel. $\{Q_i\}$ stands for the other qubits in the grid not expressly depicted that play a role in the state preparation, $U_{\text{WALA}}$, and Trotterization, $U_{\text{Trot}}$. The $P$ gate on the ancilla qubit, $Q_a$, is the state preparation gate for the ancilla, which is an $H$ gate if measuring the real part or $HS^\dagger$ gate if measuring the imaginary part. **c**, Measurements of $|\mathcal{C}(j,t)|$ obtained by adding the the real and imaginary parts in quadrature. The upper panel shows the deconfined case with $h_E = 0$, while the lower panel shows the confining case with $h_E = 2.0$. All measurements were acquired using the Trotter step $dt = 0.5$ and $\lambda = 0.25$. Distinct final string configurations shown in panel **a** are shown by color: $j = 1$ (blue), $j = 2$ (orange), $j = 3$ (green), $j = 4$ (purple). The grey region depicts the region expected to be blocked by decoherence and is bounded by the magnitude of the measured correlator after evolution of the WALA state with $\lambda = 0$. **d**, Heatmaps of $|\mathcal{C}(j,t)|$ for $h_E \in \{0, 0.3, 0.6, 0.8, 2.0\}$, after rescaling the data assuming a global depolarizing channel.

The ground-state of the toric code Hamiltonian, can be written as the superposition of all loop configurations [3, 4]. This picture suggests an intuitive framework to study confinement in terms of the tension in such strings, which connect electric excitations. In the topological phase, the electric excitations at the end points of the string are not confined. Increasing the electric field, $h_E$, past a critical value results in finite string tension and consequently confines the motion of the excitations at the ends of the string. To study the build up of this string tension directly, we measure the $X$-string two-time correlator:

$$\mathcal{C}(j,t) = \langle \psi_0 | (X_{Q_1} X_{Q_2} \ldots X_{Q_j})(t) X_{Q_1}(0) | \psi_0 \rangle, \tag{S7}$$

where $j$ is the length of the late-time string. Intuitively, one can think about Eq. (S7) as a measure of the likelihood that, given a set of charges were created on either side of qubit $Q_0$, a string stretching to $Q_j$ exists after time $t$. If there is no tension, the string will be able to grow without any penalty or oscillation. However, we expect a finite string tension to preclude the existence of long strings.

We start with a grid with an extra qubit, $Q_1$, off the edge, which is also coupled to an ancilla qubit, $Q_a$. We will consider straight strings in $\mathcal{C}$ stretching from $Q_1$ on the left towards the right (Fig. S5a). Similar to our discussion in Supplementary Information III C 1, we will not apply local field terms on qubit $Q_1$ to keep the string pinned to the edge. To measure $\mathcal{C}$, we turn to a generalized Hadamard test (Fig. S5b and Methods). The ancilla is initialized in either an eigenstate of Pauli-$X$ or Pauli-$Y$, to measure the real or imaginary part of $\mathcal{C}(j,t)$. After applying a C-NOT gate on $Q_1$, controlled by $Q_a$, the standard Trotterized unitary is applied to the full system of qubits. We choose not to apply the single qubit field terms to qubit $Q_1$, in order to fix one end of the string at the edge. Then all qubits along the string are measured in the $X$ basis. Measuring the real and imaginary parts we compute the magnitude $|\mathcal{C}|$ (all data shown in Fig. S7), we observe that for both $h_E = 0$ and 2.0, the correlator has an initial value of 1 for string length $j = 1$, while all other values of $j$ result in $|\mathcal{C}| = 0$ at $t = 0$. For $h_E = 0$, $|\mathcal{C}(1, t)|$ decays towards zero with increasing $t$, while $|\mathcal{C}(2, t)|$ increases to a value of 0.3, with $|\mathcal{C}(3, t)|$ increasing at later times to $\sim 0.1$ (Fig. S5c). In Fig. S5d, we present rescaled color plots of $|\mathcal{C}|$ for several values of $h_E$ (re-plotting $h_E = 0, 2.0$). Data for small $h_E \in \{0, 0.3\}$ is consistent with the absence of a string as the excitation is spreading over the lattice. When $h_E \in \{0.6, 0.8, 2.0\}$, the dominant signal comes from the $\mathcal{C}(1, t)$ channel across the entire time frame. This is indicative of a string, unable to stretch from its initial length, showing evidence of confinement. Indeed $\mathcal{C}(1, t)$ has strong weight for long times when $h_E$ is large, while for small $h_E$ this observable decays to zero while the longer strings sequentially pick up additional intensity as time goes on. These results confirm the onset of confining dynamics at $h_E = 0.6$. Plots and all raw and rescaled heatmap data is presented in Supplementary Information IV A.

## D. String dynamics with $\lambda = 0$

In Fig. 4 of the main text, we show plots of the string dynamics with $\lambda = 0.25$ using $\mathcal{S}_{ZZ}(t)$ and $\mathrm{Re}[\langle Z(t)Z(0)\rangle]$. We then interpret our data in terms of different deformations of the string that either move the bump in the string to the bottom side of the grid or cause it to remain on the top. However, since the string breaking parameter $\lambda \neq 0$, it is also possible that instead of undergoing one of these two deformations the string simply breaks. Indeed, our data in Fig. 5, Methods, and Supplementary Information IV B show that, when $\lambda = 0.25$, string breaking is taking place and affecting the local excitation occupation. The natural question that arises is whether this string breaking is substantially modifying the dynamical motion, and if so, how that may affect our interpretation of the data in Fig. 4.

To clarify the situation, we compare datasets with $\lambda = 0$ and $\lambda = 0.25$ (Fig. S6). Both show nearly identical behavior for the full range of $h_E$ values. This behavior is further supported by our circuit simulations (bottom panels). The hardware data agrees very well, after a global depolarization rescaling has been applied, for $\lambda = 0$ and the $\lambda = 0.25$ results from Fig. 4, reproduced here in Fig. S6c,d. Since the qualitative behavior between the two values of $\lambda$ is similar, we also plot the differences of the rescaled $\mathcal{S}_{ZZ}$ and $\mathrm{Re}[\langle Z(t)Z(0)\rangle]$ between the two different values of $\lambda$ (Fig. S6e,f). Our measurements and numerical simulations indicate a very small difference between the two $h_E$ cases, typically on the order of 5%.

Overall, the very small differences of $\mathcal{S}_{ZZ}$ and $\mathrm{Re}[\langle Z(t)Z(0)\rangle]$ between $\lambda = 0$ and $\lambda = 0.25$ support our interpretations in the main text that the dominant factor in our data is the motion of the string.

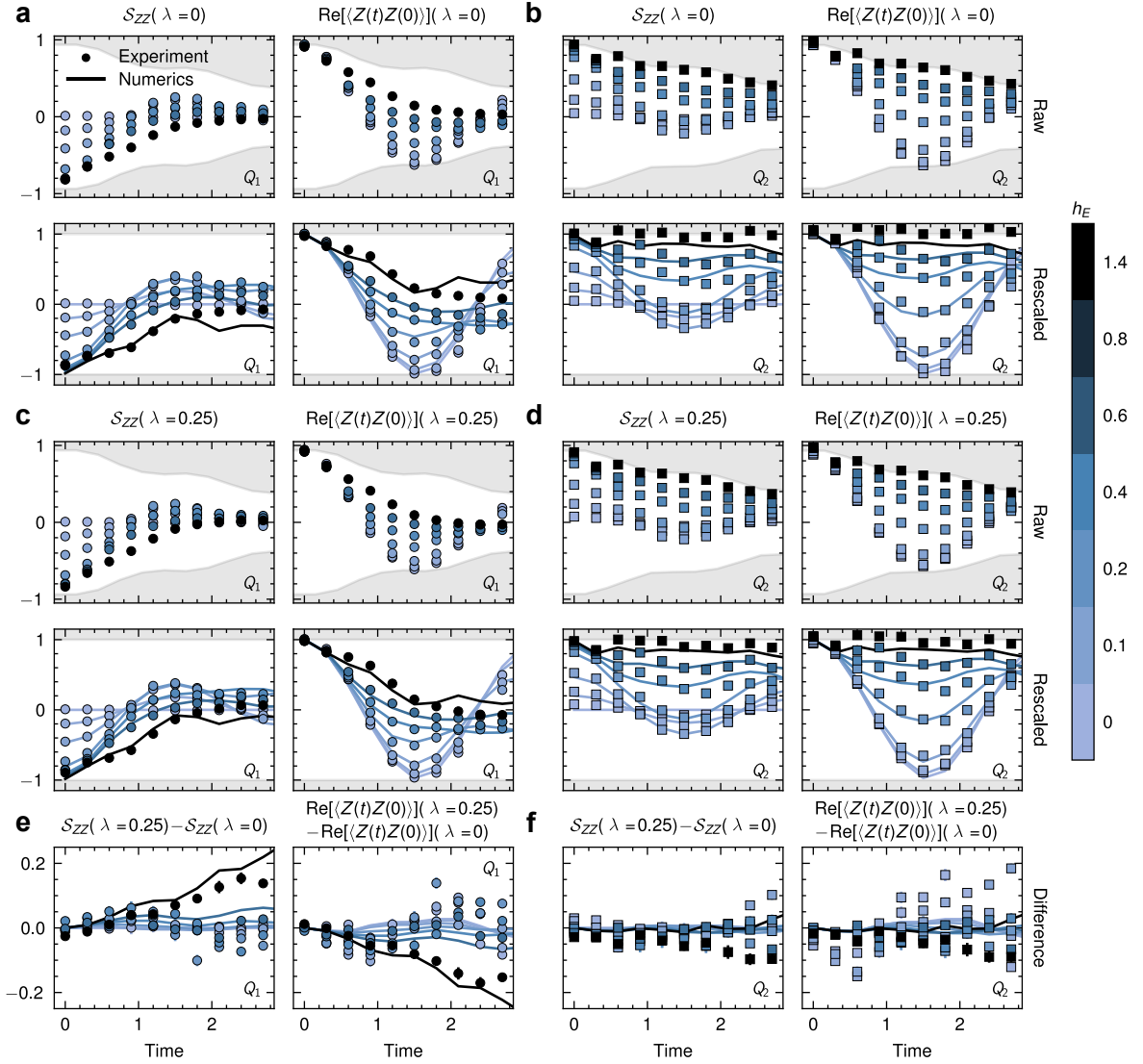

FIG. S6. **Time traces of $\mathcal{S}_{ZZ}(t)$ and $\mathrm{Re}[\langle Z(t)Z(0)\rangle]$ for $\lambda \in \{0, 0.25\}$.** Panels **a–d** all show dynamical plots of $\mathcal{S}_{ZZ}(t)$ and $\mathrm{Re}[\langle Z(t)Z(0)\rangle]$ as a function of $h_E$ (blue/black colorscale). The top two panels show raw data after only post-selecting all ancillas to be in the $|0\rangle$ state. The grey area on these plots corresponds to the region limited by decoherence and is bounded by $|\langle Z(t)Z(0)\rangle|_{\lambda=h_E=0}$. Using this curve, the bottom panels show the same data as the top panels rescaled assuming global depolarization (markers), and noiseless numerical simulations (lines). Panels **a–d** differ by the value of $\lambda$ and the qubit plotted (defined in Fig. 4). **a,** Qubit $Q_1$, $\lambda = 0$. **b,** Qubit $Q_2$, $\lambda = 0$. **c,** Qubit $Q_1$, $\lambda = 0.25$. **d,** Qubit $Q_2$, $\lambda = 0.25$. **e,f,** The difference between the rescaled $\mathcal{S}_{ZZ}$ and $\mathrm{Re}[\langle Z(t)Z(0)\rangle]$ data for $\lambda = 0.25$ and $\lambda = 0$ including the numerical simulations for qubit **e,** $Q_1$ and **f,** $Q_2$.

## IV. Additional numerical circuit simulations

### A. Real and imaginary parts of the string correlator $\mathcal{C}$

To calculate $|\mathcal{C}(j,t)|$, as shown in Fig. S5c,d, the real and imaginary parts must be collected on the device and added in quadrature. Measurements of $\mathrm{Re}[\mathcal{C}(j,t)]$ and $\mathrm{Im}[\mathcal{C}(j,t)]$ are shown in Fig. S7a,b, and compared to exact numerical simulations after a rescaling, assuming a global depolarizing model (Methods). The results of the corresponding magnitude of both mitigated and rescaled data are shown in Fig. S7c.

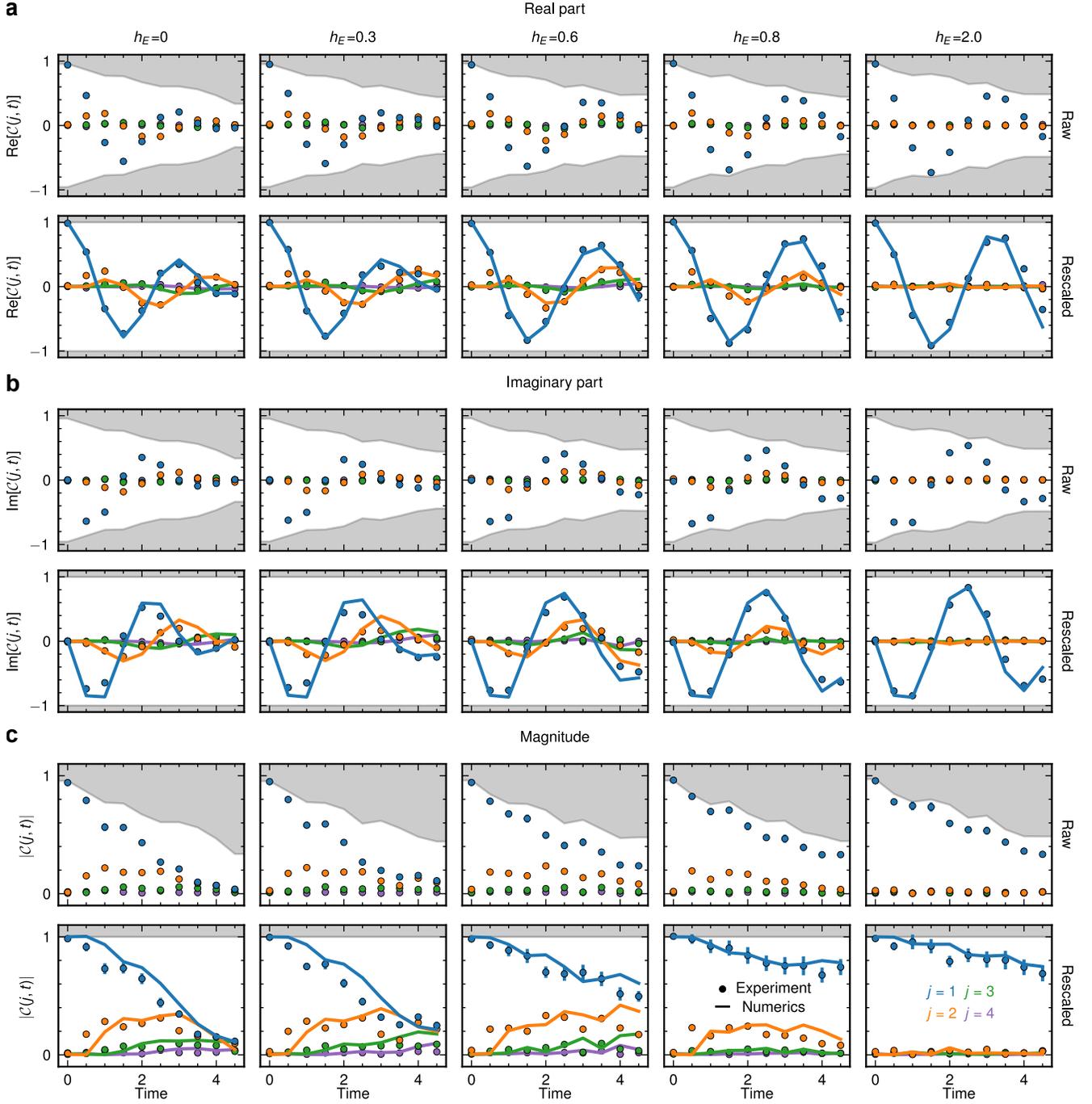

FIG. S7. **Raw and rescaled measurements of the string correlator. a**, Raw data, after post-selecting on all ancillas being in the $|0\rangle$ state, of measurements of $\mathrm{Re}[\mathcal{C}(j,t)]$, for $h_E \in \{0, 0.3, 0.6, 0.8, 2.0\}$ and constant $\lambda = 0.25$, $dt = 0.5$. The grey regions in the top panels are bounded by $\pm|\mathcal{C}(1,t)|$ under evolution of a Hamiltonian with $\lambda = 0$. The bottom row of panels shows the global depolarization rescaled values (markers) and the noiseless circuit simulations (lines). **b**, Same plots as in panel **a**, but for $\mathrm{Im}[\mathcal{C}(j,t)]$. **c**, Plots of the modulus $|\mathcal{C}(j,t)|$, extracted from the data in panels **a** and **b**.

## B. Plaquette occupation for $A_1$, $A_2$, and vacuum

To compare the values of vertex occupation, $P_{A_v}$, reported in the main text Fig. 5 to theoretical values, we calculate a rescaling parameter using the WALA state with the string excitation, evolved under a Hamiltonian with $\lambda = 0$ and $h_E = 1.4$. We then take the depolarized expectation value as the average of $P_{A_v}$ over all 12 grid vertices. This is then used to calculate a global depolarizing probability and the rescaling is performed using the fact that the expectation

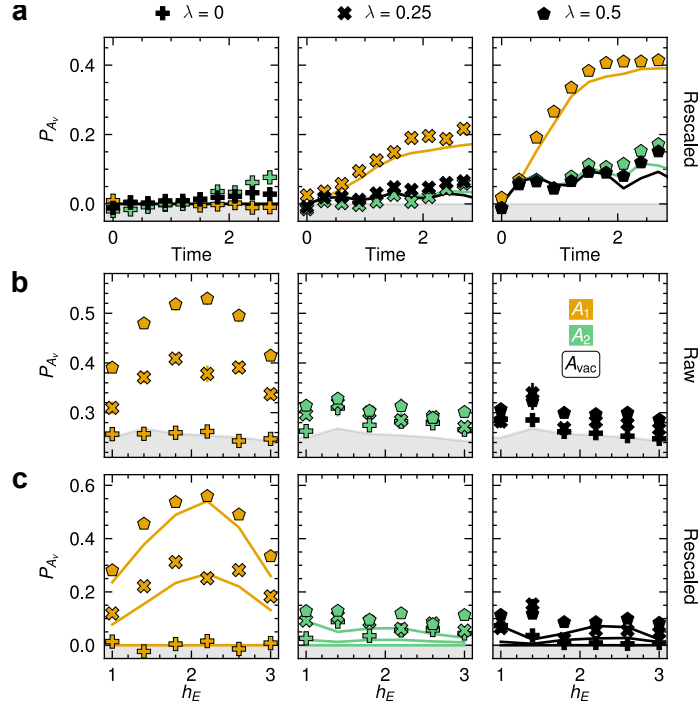

FIG. S8. **Rescaled and simulated occupation of $A_1$, $A_2$, and $A_{vac}$. a**, Vertex occupation from the main text Fig. 5c, rescaled assuming a global depolarizing model with the depolarizing probability determined from the average of $P_{A_v}$ over all vertices, having evolved the initial state with an $X$ string using a Hamiltonian with $\lambda = 0$, $h_E = 1.4$. **b**, $h_E$ dependence of the vertex occupation, $P_{A_v}$, for $\lambda = 0$ (pluses), $\lambda = 0.25$ (crosses), and $\lambda = 0.5$ (pentagons). The three panels show $P_{A_v}$ for vertices $A_1$ (gold), $A_2$ (green), and $A_{vac}$ (black). Data was acquired after ten Trotter steps of $dt = 0.2$. **c**, Rescaled and simulated values of $P_{A_v}$, rescaled using the same global depolarizing method as in **a**.

value of $P_{A_v}$ in the maximally mixed state is $1/2$ (Methods).

The comparison between theory and experiment shows excellent agreement in the time dynamics of $P_{A_v}$ (Fig. S8a). While nonzero $P_{A_v}$ is observed for $A_2$ and $A_{vac}$ when $\lambda > 0$, this can be attributed to vacuum fluctuations, since the values for $A_2$ and $A_{vac}$ are equivalent.

While the $A_1$ vertex shows a resonance near $h_E = 2.0$, our data shows that there is no such resonance for $A_2$ or $A_{vac}$ (Fig. S8b). This further supports our claim that the additional vertex occupation on $A_1$ is a consequence of string breaking. We also observe excellent agreement between numerical simulation and the rescaled resonance data for all three vertices (Fig. S8c).

## C. Trotter error

Intrinsic to Trotter evolution is the error that accumulates from non-commuting terms in the Suzuki-Trotter expansion. Generally this error is minimized by choosing smaller Trotter steps $dt$, since the Suzuki-Trotter expansion is exact in the limit $dt \to 0$. To understand the nature of these errors for the Trotterization of the LGT Hamiltonian, we can write the next term of the Baker-Campbell-Hausdorff expansion of a single Trotter step:

$$
\begin{aligned}
e^{-i(-J\sum A_v - J\sum B_p)dt}e^{-i(-\lambda \sum X - h_E \sum Z)dt} = \exp[ & -idt(-J\sum_v A_v - J\sum_p B_p - \lambda \sum_l X - h_E \sum_l Z) \\
& -idt^2 J\lambda \sum_v (Y_{v0}Z_{v1}Z_{v2}Z_{v3} + Z_{v0}Y_{v1}Z_{v2}Z_{v3} + \dots) \\
& +idt^2 Jh_E \sum_p (Y_{p0}X_{p1}X_{p2}X_{p3} + X_{p0}Y_{p1}X_{p2}X_{p3} + \dots) \\
& + O(dt^3)]
\end{aligned}
\tag{S8}
$$

with $P_{vq}$ being the Pauli matrix ($X$, $Y$, or $Z$) for qubits $q$, which make up vertex $v$. The same convention $P_{pq}$ holds for plaquettes. The next-order correction terms can thus be seen to add hopping to both electric and magnetic excitations, through the $Y$s in the Pauli strings. Therefore, we expect totally confined excitations to slowly move apart due to Trotter errors that induce spurious hopping. While these terms depend quadratically on $dt$, and thus should be suppressed for small $dt$, it is noteworthy that they depend linearly on $\lambda$ and $h_E$ for each term, respectively. Therefore, judicious choice of Hamiltonian evolution parameters is important to optimally demonstrate the deconfining/confining phases with minimal effect from Trotter error.

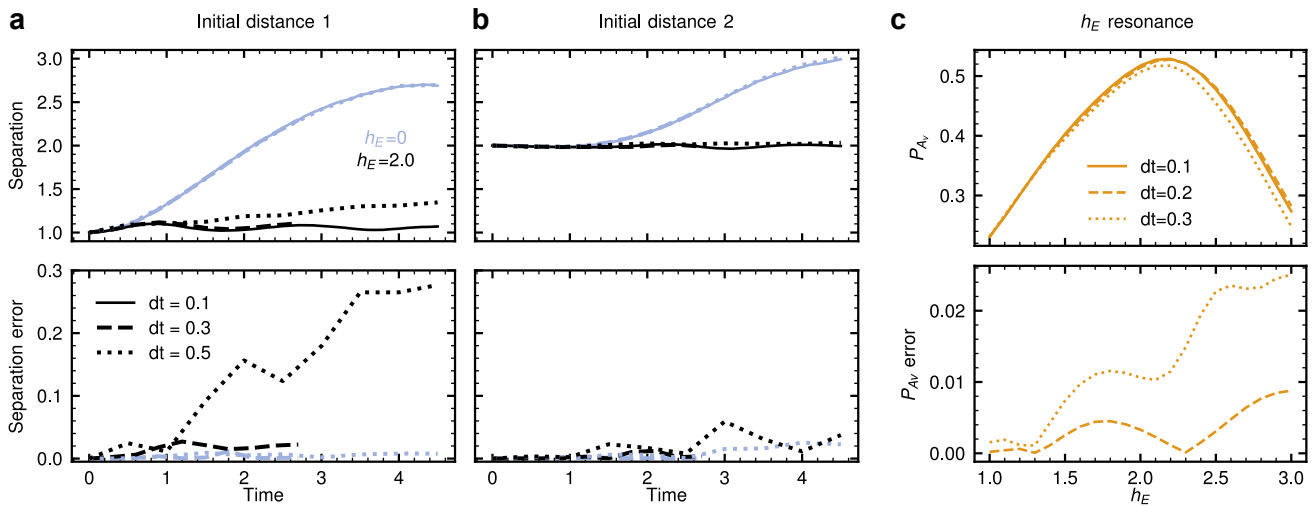

FIG. S9. **Trotter step dependence of observables. a**, Separation of excitations after creating two electric excitations with an initial distance of 1, as depicted in Fig. 3a of the main text. The top panel shows separations between excitations, simulated using $dt = 0.1$ (solid lines), $dt = 0.3$ (dashed lines), and $dt = 0.5$ (dotted line). The bottom panel estimates the Trotter error by subtracting the $dt = 0.1$ separation (negligible Trotter error) and taking the magnitude. The curves for $dt \in \{0.3, 0.5\}$ show the experimentally measurable nine Trotter steps. **b**, Same quantities as in **a**, but for an initial state with excitations separated by two sites, as shown in Fig. S3c,d. **c**, The occupancy of the $A_1$ vertex as a function of $h_E$ at time $t = 1.8$, simulated using $\lambda = 0.5$, corresponding to data presented in Fig. 5 of the main text. The top panel shows occupations simulated using $dt = 0.1$ (solid lines), $dt = 0.2$ (dashed lines), and $dt = 0.3$ (dotted line). The bottom panel estimates the Trotter error by subtracting the $dt = 0.1$ occupation (negligible Trotter error) and taking the magnitude.

To demonstrate this point, we present numerical simulations of the separation between two electric excitations for $dt \in \{0.1, 0.3, 0.5\}$ and $h_E \in \{0, 2.0\}$ (Fig. S9a). These simulations correspond to the same initial configuration considered in Fig. 3a of the main text. We see that there is practically zero Trotter error for $h_E = 0$ for all values of $dt$. However, when $h_E = 2.0$, the $dt = 0.5$ simulation shows marked departure from those using smaller Trotter steps and qualitatively loses the oscillatory behavior. As expected, when $dt$ is large the confining signatures become less clear as the excitations start to move apart from Trotter error. Therefore, for most of the results presented in the main text, the intermediate value of $dt = 0.3$ is used to maintain reasonably low Trotter error, while allowing reasonable evolution times.

The main drawback of using smaller Trotter steps is the limit imposed on the latest times that can be reached, since the device decoherence depends only on the number of cycles. Certain initial states, however, admit larger Trotter steps and thus larger effective simulations times. To this end, we have constructed a state in which the separation between charges is more robust against Trotter error (Supplementary Information III B). By starting excitations a distance of two sites apart, we allow them to both hop towards and away from each other. Since all hops are equally likely in the leading order Trotter error, a cancellation occurs when looking at the average separation between excitations. Indeed, even for $h_E = 2.0$, very small Trotter error is seen from the state with initial separation of two (Fig. S9b).

Lastly, we consider the Trotter errors when measuring the $h_E$ dependence in Fig. 5d of the main text. To show the peak at $h_E = 2.0$, we need to go to larger $h_E$ than used in the other results of the paper. Since we expect the Trotter errors to scale linearly with $h_E$, it is important to make sure we are in a low-error regime for the entire $h_E$ range. To compare, we simulate the $h_E$ dependence for $dt \in \{0.1, 0.2, 0.3\}$ and compare the occupation of site $A_1$ at the mutually achievable time of $t = 1.8$ (Fig. S9c). Indeed, we see that as $h_E$ increases, errors start to appear for $dt = 0.3$, which had given minimal Trotter error for the other results of this work. Therefore, in the main text we present experimental measurements for $dt = 0.2$ to ensure Trotter error is negligible for the full $h_E$ range.

# V. Tensor network simulations

In this section we obtain an estimate of the computational resources that are required to classically simulate the dynamics studied in the main text using tensor network methods. Even though the system we consider is two-dimensional, one-dimensional matrix-product state (MPS) simulations remain one of the state-of-the-art methods for time-evolving quantum many-body systems. Thus, in the following, we also use MPS to simulate time evolution, using the TeNPy Python library [5].

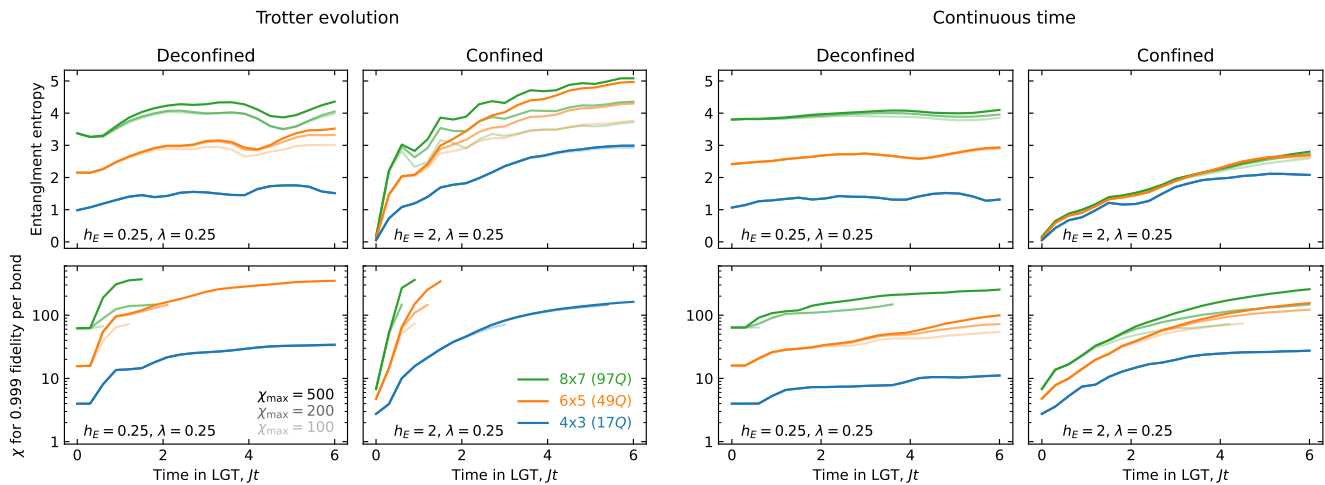

FIG. S10. **Entanglement entropy and required bond dimension in the tensor network simulation of the time evolution of an initial string excitation as in Fig. 4 in the main text.** To investigate the cost of classically simulating the experiment with MPS, we plot the maximum of the entanglement entropies of all bonds of the MPS for a given time (top row), as well as the bond dimension required to achieve 0.999 fidelity per bond (bottom row) relative to the state computed using the indicated $\chi_{max}$, i.e. the minimal number of Schmidt coefficients such that the sum of their squares is at least 0.999 for all bonds, where the Schmidt coefficients are computed using the indicated $\chi_{max}$. To avoid saturation, we end the simulation when the required bond dimension reaches 75% of $\chi_{max}$. The panels labeled "Trotter evolution" show the results for creating the string excitation on top of the WALA initial state and simulating the Trotterized time evolution with a step size $dt = 0.3$ as in the experiment. The panels labeled "Continuous time" show the results for creating the string excitation on top of the ground state obtained from DMRG and simulating the exact time evolution. The columns show either the deconfined phase ($h_E = 0.25$, $\lambda = 0.25$) or the confined phase ($h_E = 2.00$, $\lambda = 0.25$). The different colors show different system sizes, defined by the number of vertex sites for electric charges. Different shades of a color correspond to simulations with different bond dimensions, $\chi_{max}$, of the same system size. The legend in the lower-right panel indicates the number of qubits without the two extra qubits needed to pin the ends of the string excitation, as these two qubits remain unentangled throughout the time evolution and could be removed from the simulation.

In particular, we consider a setup corresponding to the experiment in Fig. 4 in the main text, where a string is stretched across the system with a single bump in the middle. To simulate the system with MPS, we take the layout of the qubits as shown in Fig. 4a (without any ancilla qubits) and order them column by column to obtain a one-dimensional geometry suitable for MPS simulations. The MPS simulation is efficient if the bond dimension remains small during the time evolution. Because the required bond dimension at a given bond scales exponentially with the entanglement entropy across that cut, we plot the maximum entanglement entropy over all bonds of the MPS in the top panels of Fig. S10. In the lower panels of the same figure, we directly show the required bond dimension to achieve 99.9% fidelity for each bond relative to the state computed with the indicated $\chi_{max}$.

We consider two setups for simulating the time evolution, labeled "Trotter evolution" and "Continuous time," respectively, in the figure. For the former, we mimic the experiment by creating the string excitation on top of the WALA state and simulating the Trotterized time evolution with step size $dt = 0.3$. For the latter, we simulate the continuous dynamics by creating the string excitation on top of the ground state obtained from DMRG [5, 6] and approximate the exact time evolution. To obtain the results for the continuous time evolution, we use the time-dependent variational principle (TDVP) [5, 7, 8]. We first use a two-site variant of TDVP to grow the bond dimension starting from the initial state until it is saturated, before switching to the usual single-site variant of TDVP. We use a step size $\delta t = 0.03$, and the entanglement entropy and the required bond dimension are calculated every 10 steps, corresponding to $dt = 0.3$. For the first-order Trotterized time evolution, we can apply the terms involving only the onsite field terms as onsite operators to the MPS, and evolve the state with the plaquette and vertex terms using the same TDVP algorithm as above [5, 7, 8] with $\delta t = 0.03$ for 10 steps until we reach the duration of the Trotter step $dt = 0.3$.

The first and third columns of Fig. S10 show the evolution of the entanglement entropy and the required bond dimension in the deconfined case for $h_E = 0.25$ and $\lambda = 0.25$. The entanglement entropy increases linearly with the system size, and the required bond dimension increases exponentially, reflecting that the one-dimensional MPS ansatz with fixed bond dimension cannot capture area law entanglement in two dimensions. These quantities also grow moderately in time.

The second and fourth columns show the same quantities in the confined case for $h_E = 2.0$ and $\lambda = 0.25$. While the initial entanglement entropy and required bond dimension are much smaller since the state is closer to a product state, both quantities quickly grow in time. This is expected since in the confined case the energy of the string is extensive in its length, and so we inject much more energy into the system than in the deconfined phase. The simulation of the experimental circuit ("Trotter evolution") shows an even faster growth of the entanglement entropy and required bond dimension, as both the Trotter approximation of the time evolution and the WALA state not being an exact eigenstate lead to the creation of further excitations on top of the initial state, which increases the entanglement. In the confined case, the entanglement entropy and required bond dimension show relatively little system-size dependence in the continuous time case.

In the deconfined case, simulating the continuous time evolution of the $8 \times 7$ system with $\chi_{\max} = 500$ took 26 hours on a `c3d-standard-360` node [9] on Google Cloud with 180 vCPUs (hyperthreading disabled) and 1.4 TB of available RAM. The algorithm scales as $\chi^3$, and Fig. S10 indicates that the required bond dimension scales exponentially with the system size. Therefore, simulating the string dynamics in the deconfined regime will quickly become prohibitively expensive as we scale up the system size. While using more optimized simulation methods can probably push the simulation capabilities of classical computers to somewhat larger systems and later times than considered here, the area law of the entanglement entropy ultimately limits the classical simulation. Therefore, these considerations indicate a route where eventually larger systems and longer timescales will be beyond the capabilities of classical tensor network simulations.

---